\documentclass[10pt, aps, pre, reprint, onecolumn, superscriptaddress, titlepage, amsmath,amssymb]{revtex4-2}

\pdfoutput=1

\usepackage{amssymb,amsmath}

\usepackage[utf8]{inputenc}
\usepackage{csquotes} 
\usepackage{graphicx}
\usepackage{amsthm}
\usepackage{cancel}
\usepackage{color}
\usepackage{bbold}
\usepackage{wrapfig}
\usepackage{booktabs}
\usepackage{float}
\usepackage{enumitem}
\usepackage{hyperref}
\usepackage{tikz}
\usepackage{comment}
\usepackage{longtable}
\usepackage{multirow}
\usepackage{url}
\usepackage{siunitx}

\usepackage{xcolor}

\usepackage{comment}

\usepackage[american]{babel}

\usepackage{hyperref}

\usepackage[normalem]{ulem} 
\newcommand{\ts}{\,\Delta t_\mathrm{stop}}

\newcommand{\lav}{\,\langle \ell \rangle}
\newcommand{\oav}{\,\langle O\rangle}

\newcommand{\kmh}{\,\mathrm{km}/\mathrm{h}}
\newcommand*\Lrel{%
  \mathchoice
    {\frac{L_\mathrm{driv}}{L_\mathrm{req}}} 
    {L_\mathrm{driv}/L_\mathrm{req}} 
    {L_\mathrm{driv}/L_\mathrm{req}} 
    {L_\mathrm{driv}/L_\mathrm{req}} 
}

\usepackage{mathtools}
\usepackage{commath}
\usepackage{nicefrac}
\usepackage{todonotes}

\begin{document}

\date{\today}
\author{Charlotte Lotze} 
\thanks{These authors contributed equally.}
\affiliation{Chair of Network Dynamics, Center for Advancing Electronics Dresden (cfaed) and Institute of Theoretical Physics, Technische Universität Dresden, 01062 Dresden, Germany}

\author{Philip Marszal} 
\thanks{These authors contributed equally.}
\affiliation{Chair of Network Dynamics, Center for Advancing Electronics Dresden (cfaed) and Institute of Theoretical Physics, Technische Universität Dresden, 01062 Dresden, Germany}

\author{Felix Jung} 
\thanks{These authors contributed equally.}
\affiliation{Chair of Network Dynamics, Center for Advancing Electronics Dresden (cfaed) and Institute of Theoretical Physics, Technische Universität Dresden, 01062 Dresden, Germany}

\author{Debsankha Manik}
\affiliation{Chair of Network Dynamics, Center for Advancing Electronics Dresden (cfaed) and Institute of Theoretical Physics, Technische Universität Dresden, 01062 Dresden, Germany}

\author{Marc Timme} 
\thanks{These authors contributed equally. }
\affiliation{Chair of Network Dynamics, Center for Advancing Electronics Dresden (cfaed) and Institute of Theoretical Physics, Technische Universität Dresden, 01062 Dresden, Germany}
\affiliation{Lakeside Labs, 9020 Klagenfurt, Austria}

\author{Malte Schröder} 
\thanks{These authors contributed equally. }
\affiliation{Chair of Network Dynamics, Center for Advancing Electronics Dresden (cfaed) and Institute of Theoretical Physics, Technische Universität Dresden, 01062 Dresden, Germany}

\title{Identifying the threshold to sustainable ridepooling}

\begin{abstract}
Ridepooling combines trips of multiple passengers in the same vehicle and may thereby provide a more sustainable option than transport by private cars. The efficiency and sustainability of ridepooling is typically quantified by key performance indicators such as the average vehicle occupancy or the total distance driven by all ridepooling vehicles relative to individual transport. However, even if the average occupancy is high and rides are shared, ridepooling services may increase the total distance driven due to additional detours and deadheading. Moreover, these key performance indicators are difficult to predict without large-scale simulations or actual ridepooling operation. 
Here, we propose a dimensionless parameter to estimate the sustainability of ridepooling by quantifying the \textit{load} on a ridepooling service, relating characteristic timescales of demand and supply. The load bounds the relative distance driven and uniquely marks the break-even point above which the total distance driven by all vehicles of a ridepooling service falls below that of motorized individual transport. Detailed event-based simulations and a comparison with empirical observations from a ridepooling pilot project in a rural area of Germany validate the theoretical prediction. Importantly, the load follows directly from a small set of aggregate parameters of the service setting and is thus predictable \textit{a priori}. The load may thus complement standard key performance indicators and simplify planning, operation and evaluation of ridepooling services.
\end{abstract}

\maketitle

\section{Introduction}
Ridepooling promises to reduce the number of vehicles on the road as well as the total distance driven \citep{santi2014, alonso-mora2017} by combining trips of multiple passengers with similar origin and destination in the same vehicle. In principle, ridepooling services are particularly efficient in high-demand urban mobility with many similar trips \citep{santi2014, tachet2017, zwick2021ride, storch2021}, but comparable on-demand services have also been tested in rural areas to supplement sparse conventional public transport \citep{ecobus, burstlein2021, kaddoura2021}. However, it is difficult to predict \textit{a priori} under which conditions such services are ecologically more sustainable than private cars. 
For example, even in densely populated urban areas with a high potential for sharing trips, on-demand mobility services may increase the total number of vehicles on the road, leading to congestion and lower average travel speeds \citep{erhardt2019, henao2019, erhardt2021, diao2021}. Regulating and planning sustainable ridepooling services thus requires (i) a way to quantify the ecological sustainability of ridepooling services and (ii) a method to estimate the expected sustainability already before a service is running. 

Various key performance indicators are used to quantify different aspects of the operational efficiency of ridepooling services. The \textit{relative distance driven} with respect to individual motorized mobility is a simple key performance indicator that roughly captures the ecological sustainability in terms of the carbon emissions of a ridepooling fleet compared to equivalent trips in private cars \citep{liebchen2021}. Alternatively, research analyses and current regulations often rely on the average \textit{occupancy} of vehicles as a key performance indicator, computed as the total passenger distance traveled divided by total vehicle distance driven, for example in New York City \citep{wolfe2018, honan2019} and Germany (\citep{pbefg2021}, § 50 Absatz 3). Although the occupancy quantifies the utilization of the ridepooling vehicles, its interpretation in terms of ecological sustainability is not straightforward because of detours from pooling trips of multiple passengers. Despite transporting more than one passenger at a time, ridepooling services may thus fail to be more sustainable than private cars.

Moreover, while these key performance indicators are easy to evaluate for services already in operation, it is difficult to estimate them before running the service. For example, the relative distance driven by a ridepooling service may depend on differences across service locations, service parameters, and demand settings. In particular, no method exists to predict under which conditions ridepooling services become more sustainable than private cars, in the sense that the total distance driven by a ridepooling service breaks even with the total direct distance of requested trips. 

Here, we propose a dimensionless parameter to characterize the sustainability of ridepooling by quantifying the \emph{load} on a ridepooling (or ridehailing) service based on the balance of characteristic demand and supply timescales \citep{molkenthin2020}. In contrast to standard key performance indicators, the load may be computed without direct observation or simulation of a ridepooling service based on a small set of aggregate system parameters. It may thus complement other key performance indicators by facilitating the comparison of services across settings and independent of operational details \citep{molkenthin2020, zech2022}. 
Moreover, the load is directly related to the relative distance driven and uniquely indicates the break-even point of the total distance driven compared to individual mobility. We illustrate this relationship in detailed ridepooling simulations and demonstrate its robustness across different settings and empirical data from a ridepooling pilot project in rural Germany.

\section{Research question and context}
The potential of ridepooling to increase efficiency and sustainability of transport compared to individual mobility is analyzed and highlighted by several studies. For instance, \citet{santi2014} exemplified this potential by showing that even pooling rides of only two passengers could reduce the number of taxi trips in Manhattan (New York City) by more than $40\,\%$. Moreover, ridepooling services require considerably fewer vehicles than non-shared taxi services while still serving most users with little delay \citep{agatz2012, vazifeh2018, spieser2014, zwick2021agent}. For example, Manhattan would require $30\,\%$ fewer vehicles \citep{vazifeh2018} when pooling rides of two passengers and could even achieve more than a $75\,\%$ reduction with larger vehicles \citep{alonso-mora2017}. In Singapore, a reduction of the number of vehicles by approximately $66\,\%$ may be possible \citep{spieser2014}. In this way, ridepooling promises to reduce both the total distance driven and the fleet size, resulting in significantly lower energy consumption and carbon emissions and thus higher ecological sustainability. Barann et al.~estimate potential savings in New York City alone as about three million kilometers of travel distance and more than 500 tons of carbon emissions per week \citep{barann2017}.

Various different measures to quantify the efficiency of ridepooling have been proposed. For example, Santi et al.~have mathematically formalized the pooling of ride requests as a graph-theoretical matching problem to quantify the \textit{shareability} of rides. They define the fraction of individual trips that are shareable with at least one other trip without exceeding a maximum delay as a characteristic measure of the theoretical potential of ridepooling \citep{santi2014}. Building on these results, \citet{tachet2017} reveal universal scaling of the shareability with the total number of requests across different cities, supporting the transfer of insights across specific cities and settings. Moreover, this formalism enables predictions of the operational parameters such as the required fleet size for ridehailing and ridepooling services discussed above \citep{vazifeh2018}. A similar approach to quantify the efficiency of ridepooling services includes the spatial overlap of requests and vehicle routes derived solely from geometric and dimensional arguments by \citet{herminghaus2019}. Molkenthin et al.~define the efficiency of a ridepooling service based on its collective dynamics by considering the average number of requests in the system at any given time \citep{molkenthin2020}. This number of scheduled requests is directly related to the service time of a request and characterizes both the throughput from the perspective of the ridepooling operator and the service quality from the perspective of the users. A similar measure in terms of the number of scheduled stops is used by \citet{manik2020}. Other approaches to quantify the efficiency include, for example, the average vehicle occupancy \citep{liebchen2021, pbefg2021} or the acceptance rate of requests \citep{fielbaum2021b, alonso-mora2017, deruijter2020, deRuijter2023ride}. All these efficiency measures describe different aspects of the efficiency of ridepooling services, from social costs for the users \cite{erhardt2019, storch2021}, economical feasibility of the service from the point of view of the operator \citet{vazifeh2018, herminghaus2019}, or environmental benefits in terms of sustainability \cite{barann2017, muhle2023}. 

The ecological sustainability is a particularly relevant aspect of ridepooling efficiency. The \textit{total distance driven} approximately quantifies the total energy consumption and emissions. This measure is referred to by a variety of names such as “total vehicle movement distance” \citep{deruijter2020}, “vehicle kilometers traveled” \citep{beojone2021}, “vehicle miles traveled” \citep{engelhardt2019, fagnant2018, ruch2020}, “distance driven” \citep{engelhardt2020}, “vehicle hours traveled \citep{fielbaum2021b}, “vehicle travel distance” \citep{ma2013a}, or “total route length” \citep{lotze2022}. It is widely used to evaluate the efficiency and sustainability of ridepooling, for example in terms of the relative distance driven of a ridepooling service with respect to private cars, and applied to compare different algorithms for implementing the matching of requests and the routing of vehicles in a ridepooling service \citep{alonso-mora2017, fielbaum2021b, engelhardt2020, ma2013a}.

However, whether ridepooling reduces the distance driven and is ecologically more sustainable than individual motorized mobility depends on the specific setting of the ridepooling system. While most studies emphasize the high potential of ridepooling to reduce total distance driven and emissions \citep{santi2014, vazifeh2018, spieser2014, alonso-mora2017, herminghaus2019, lotze2022}, others hint at the possibility that ridepooling services may increase congestion and distance driven \citep{beojone2021, engelhardt2019}. Moreover, a case study in San Francisco, CA, shows that competition with public transport and pull-effects of these services may actually increase the total number of trips and cars on the road \citep{erhardt2019, erhardt2021}. 

To ensure sustainability of ridepooling services, it is essential to understand in which settings ridepooling is more sustainable than individual motorized mobility. Several studies estimate the potential efficiency or inefficiency of ridepooling services in specific cities across the globe such as New York City, NY \citep{santi2014, barann2017}, Munich \citep{engelhardt2019}, or Shenzhen \citep{beojone2021}. But a general understanding of these conditions is still missing. In the following, we introduce the load of a ridepooling service as an easy-to-compute dimensionless indicator that characterizes relevant aspects of the ridepooling setting and marks the break-even point where ridepooling begins to reduce the total distance driven. We show that the load provides an upper bound for the relative distance driven compared to individual mobility without requiring detailed empirical data or simulations of ridepooling in a specific setting.

\newpage

\section{Methods and Data}

Ridepooling extends standard ridehailing services by combining trips of multiple passengers into the same vehicle to increase utilization of vehicles and reduce total distance driven. In general, a ridepooling service works as follows: A user makes a request for transportation from a specific origin to a specific destination. The ridepooling service makes an offer for transportation including the (approximate) trip duration and arrival time to the user, depending on the available vehicles and their currently planned trips. If no suitable vehicle is available, the service may not make an offer and reject the request. Conversely, if the user is not satisfied with the offer, they may decline the offer. 
If the user accepts, the appropriate vehicle of the fleet is (re)routed to pick up the user at their origin and delivers them to their destination. The selection of possible offers and the vehicle assignment and routing are done such that they optimize (some measure of) the service quality and efficiency of the ridepooling fleet, balancing between minimizing waiting times, detours, or delays for passengers in the same vehicle while aiming for a high utilization per vehicle and serving as many users as possible. Over time, the planned routes of vehicles evolve as more users make requests and vehicles serve the users along their planned routes. 

In the following, we describe the event-based simulations including different extensions and the empirical data analyzed in this manuscript.

\subsection{Ridepooling model}
We derive our results in a simplified theoretical model of ridepooling under constant conditions that allows for a simple mathematical treatment.\\

\paragraph{Spatial setting} 
We model ridepooling on a 2D unit square with continuous positions $(x,y)$ with $x,y \in [0,1)$ with periodic boundary conditions, i.e., a vehicle leaving the area to the right (top) enters again on the left (bottom). This simplification makes the system homogeneous and avoids boundary effects, enabling simpler mathematical treatment. We relax these conditions to demonstrate the robustness of our analysis in various additional simulations explained below.\\

\paragraph{Request distribution} 
Requests appear with a constant rate $\lambda$ following a Poisson process in time, such that the time between two consecutive requests is exponentially distributed with expected time $1/\lambda$. The origin of each request is chosen uniformly randomly in the unit square, the destination uniformly randomly in a disk of radius $\ell_\mathrm{max} = 1/2$ around the origin, resulting in an average trip distance $\lav = 1/3$. This again ensures a homogeneous and symmetric request distribution and is relaxed in various additional simulations explained below.\\

\paragraph{Vehicle fleet} 
The fleet of the ridepooling service consists of $B$ vehicles. The vehicles drive along their planned route with constant velocity $v = 1$, without loss of generality, by measuring time and space in appropriate units. We model picking up or dropping off passengers as instantaneous, such that vehicles do not stop or slow down at their scheduled stops. When a vehicle has no further stops scheduled, it becomes idle and waits at its current location until it is assigned another request. Additionally, we assume that all vehicles have sufficient capacity to serve any number of requests. This simplification avoids overload of the system where the system cannot serve all incoming requests, again enabling an initially simpler theoretical description. \\

\paragraph{Assignment and routing algorithm} 
When a user requests a trip, they are immediately assigned to one of the $B$ ridepooling vehicles and the origin and destination of the request are added to the planned route of that vehicle. We do not post-process previously accepted requests, for example by assigning them to a different vehicle. However, other requests may be delayed when a new request is accepted. To find a suitable vehicle and routing plan, we employ a simple greedy algorithm. For each possible insertion of the origin and destination into the planned route of a vehicle, we compute the time until the vehicle would become idle, i.e., the total time until the vehicle finishes all scheduled requests. For each vehicle, we select the insertion that minimizes this time, resulting in the vehicle serving all its assigned requests as fast as possible, thereby automatically also keeping travel times small. If multiple insertions result in the same time until the vehicle becomes idle, we select the insertion that results in the fastest arrival of the new request. Out of these single-vehicle-optimal insertions, we then assign the request to the vehicle that becomes idle the fastest. This assignment algorithm thus optimizes the overall service time from the point of view of the vehicle fleet. While it does not explicitly prevent long delays for individual users, this typically also results in fast service for users.

\subsection{Ridepooling simulation}
We initialize the simulation with all $B$ vehicles distributed uniformly randomly across the simulation space and initially idle without any scheduled requests. Starting from this state, we simulate the dynamics in an event-based framework: We first compute the time until the next event occurs, e.g., a new request appears, or a vehicle reaches a scheduled stop. Then, we advance the time and update the state of the system in terms of the positions, occupancy, and planned routes of all vehicles to this new time. We repeat the process until a total time of $T_\mathrm{sim} = 300$ has passed or at least $1000$ requests per vehicle have occurred. 

During the simulation, we record all events (new requests, vehicles reaching stops etc.). This data allows to fully reconstruct the state of the system, enabling us to measure any observable of the system at any point in time. Disregarding an initial equilibration period of $100$ time steps, we compute (i) the total requested distance $L_\mathrm{req}$ as the sum of the direct shortest paths between the origin and destination over all requests, i.e., the total distance of the trips if driven by private car, (ii) the time-averaged occupancy of each vehicle $\oav$, and (iii) the total distance $L_\mathrm{driv}$ driven by all vehicles. Due to the long simulation time, stochastic fluctuations of the measured quantities are small, and error bars are not visible in the figures. 

We repeat these simulations with various request rates $\lambda$, representing different demand settings, and various fleet sizes $B$, representing the simplest parameter that can be directly affected by the service provider. 

\subsubsection{Ridepooling simulation extensions}
To demonstrate the robustness of our results, we conduct additional simulations in different settings and with different constraints, relaxing the simplifying assumptions made above, including stop times, capacity limits, and more complex request distributions. All simulations use the same parameters as the basic model unless changes are specified below.\\[-3mm]

\paragraph{Stop time} ---
We conduct simulations with an explicit stop time, resulting in a delay at each stop, to model the time vehicles need to decelerate, pick up or drop off passengers, and accelerate again. We simulate stop times $\ts \in \{0,0.007,0.014,0.021,0.028\}$. These stop times correspond to $\ts\in \{0,20,40,60,80\}~\text{seconds}$ when the area of the unit square represents $64~\text{km}^2$ and vehicle velocities are $10\kmh$, approximately matching conditions during rush hour in Manhattan, NY.\\[-3mm]

\paragraph{Limited capacity} ---
We conduct simulations with an explicit capacity limit $c \in \{4, 8, 16, 24\}$ for each vehicle. For simplicity, we do not reject any requests and follow the same assignment rules as in the basic model. Consequently, the system overloads when the request rate exceeds the total capacity of the system.\\[-3mm]

\paragraph{Inhomogeneous request distribution} ---
We conduct simulations with an inhomogeneous request distribution and fixed boundary conditions. We select symmetric requests $(\mathbf{o},\mathbf{d})$ with uncorrelated origin and destination from a distribution $P(\mathbf{o},\mathbf{d}) = P_o(\mathbf{o})\,P_d(\mathbf{d})$, with the same probability distribution $P_o(x,y) = P_d(x,y) \propto \mathrm{exp}\left(-\frac{\left(x-\mu_x\right)^2}{2\sigma^2}-\frac{\left(y-\mu_y\right)^2}{2\sigma^2}\right)$ restricted to the unit square with $\mu_x = \mu_y = 1/2$ for both origin and destination. We vary the heterogeneity by changing the standard deviation $\sigma \in \{0.1,0.2,0.3\}$. Small values of $\sigma$ make the system more heterogeneous with a highly concentrated demand. \\
\noindent  We start with all vehicles in the center $(\mu_x,\mu_y)$. We do not explicitly rebalance vehicles during simulations. However, the assignment algorithm equalizes planned service times across transporters, thereby automatically relocating idle transporters by preferential assignment of requests in most settings.\\[-3mm]

\paragraph{Asymmetric request distribution} ---
We conduct simulations with an inhomogeneous and asymmetric request distribution and fixed boundary conditions. We select requests $(\mathbf{o},\mathbf{d})$ with uncorrelated origin $\mathbf{o}$ and destination $\mathbf{d}$ from a distribution $P(\mathbf{o},\mathbf{d}) = P_o(\mathbf{o})\,P_d(\mathbf{d})$, with different probability distributions $P_o(x,y) \propto  \mathrm{exp}\left(-\frac{\left(x-\mu_{o,x}\right)^2}{2\sigma^2}-\frac{\left(y-\mu_{o,y}\right)^2}{2\sigma^2}\right)$ with $\mu_{o,x} = \mu_{o,y} = 0.3$ for the origin and $P_d(x,y) \propto \mathrm{exp}\left(-\frac{\left(x-\mu_{d,x}\right)^2}{2\sigma^2}-\frac{\left(y-\mu_{d,y}\right)^2}{2\sigma^2}\right)$ with $\mu_{d,x} = \mu_{d,y} = 0.7$ for the destination. We vary the heterogeneity and the directedness of the requests by changing the standard deviation $\sigma \in \{0.1,0.2,0.3\}$ of both distributions. Small $\sigma$ make the system more heterogeneous and the requests more directed.\\
\noindent We start with all buses in the center of origins $(\mu_{o,x},\mu_{o,y}) = (0.3, 0.3)$. As in the previous setting, we do not explicitly rebalance vehicles.

\newpage

\subsection{Empirical demand-responsive ridepooling data}
\begin{figure}[ht]
    \includegraphics{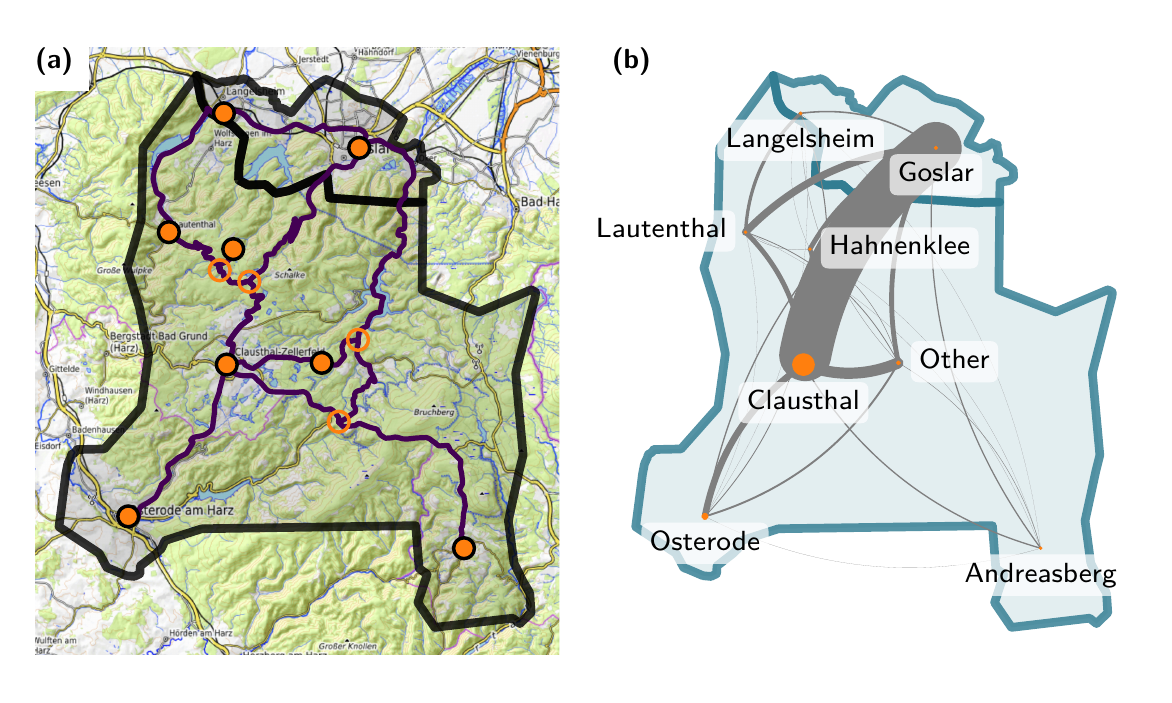}
    \caption{
        \textbf{Experimental ridepooling data.} 
        \textbf{a} Service area of the pilot project operation in Harz mountains, Lower Saxony, Germany. Requests with both origin and destination inside the smaller sub-area in the north were not eligible for service to avoid direct competition with existing public transport lines. Model simulations run on a coarse-grained network (orange nodes and highlighted streets). \textbf{b} Network of requests filed between major request centers in the service area (compare orange nodes in panel a). The width of the edges represents the number of requests between the respective regions, the diameter of the nodes represents the number of requests within the respective region. (illustrated with \citep{OSM, erhardt2021a}).
    }
    \label{fig:FIG3_ecobus_data}
\end{figure}

We compare our theoretical and simulation results to empirical data from a pilot operation of an on-demand ridepooling service that operated in rural areas of Germany (as part of the scientific project \textit{EcoBus}, see \citep{ecobus}). During the trial period, it was operated in a rural mountainous region of Harz (Lower Saxony, Germany) near the towns of Goslar and Clausthal. The exact area covered by the service is shown in Fig.~\ref{fig:FIG3_ecobus_data}a. The service was operated continuously from August 11, 2018, until February 28, 2019 \citep{ecobus}, offering transportation between 6 am and 8 pm during the weekdays and until 2 am on Friday and Saturday nights. The vehicle fleet consisted of up to 10 passenger vans with space for $8$ passengers per vehicle. Some of the vehicles additionally included the option to transport a wheelchair.

To use the service, users would need to request a ride using either a smartphone application available for iOS and Android, a web application, or by making a phone call. Arbitrary locations within the service area could be specified as origin and destination locations (except for trips within the region around Goslar, see Fig.~\ref{fig:FIG3_ecobus_data}a). In addition, users could not only book a ride with immediate pick-up, but also schedule a later time on the same day after which a pick-up should occur (\textit{pre-booking}). In the event of a successful booking, the user would receive a time frame during which their pick-up was scheduled to occur. Shortly before the pick-up, the user would receive a notification specifying the exact time of the pick-up. Trips could be canceled at the discretion of the user at any time free of charge; payment of the transport fare would take place in cash when boarding the vehicle. In December 2018, the pre-booking time was restricted to one hour in advance and the number of active bookings per user was restricted to one in order to give the system more freedom to flexibly pool and reduce misuse.

The service was integrated into the standard local transport fare system so that users would pay the same fare as when using a scheduled line service. While transportation from arbitrary locations and in smaller vehicles made the service higher-value compared to scheduled services, users had to endure the potentially large detours that the transporters would take to facilitate pooling of rides. Nevertheless, the demand was very high, leading to the rejection of most of the requests for transportation. In total, the pilot service served $38\,777$ out of $159\,223$ requests. Out of the rejected requests, $77.5\,\%$ were canceled by the users themselves, potentially due to unfavorable service (time) conditions, $19.8\,\%$ were rejected by the system upon request and $2.7\,\%$ were deleted by drivers, e.g., when users were not present at the pick-up location. Demand-side aspects of this project have been analyzed in \cite{kersting2021young, sorensen2021much}.

\subsubsection{Model simulation}
We compare the available empirical data to a model simulation in a similar setting. Using a Gaussian mixture model, we cluster the origin and destination locations of all requests into one out of eight sets. Seven sets correspond to major settlements in the service area, one set includes all other requests. For each of these sets, we compute the centroid (solid orange circles in Fig.~\ref{fig:FIG3_ecobus_data}a) as the average of all locations assigned to this set. From this clustering, we generate a coarse-grained request graph, counting the total number of requests between these sets (Fig.~\ref{fig:FIG3_ecobus_data}b).

We model the service area by a street network of $12$ discrete locations. Eight of these locations represent the centroids of the coarse-grained origin-destination demand defined above, (seven major settlements, one for all other requests, solid orange circles in Fig.~\ref{fig:FIG3_ecobus_data}a). The remaining four locations represent street intersections along the shortest paths between the centroids and are not origin or destination of requests (open orange circles in Fig.~\ref{fig:FIG3_ecobus_data}a). The travel times between the nodes in the network are approximated using Open Source Routing Machine \cite{osrm}. As the travel times slightly depend on the direction of travel, we compute them for both directions and average the results. Figure \ref{fig:FIG3_ecobus_data}a shows the computed route geometries in deep purple. For simplicity, the figure contains only routes for a single direction of travel.

For the model simulations, we assume requests distributed as $P(\mathbf{o},\mathbf{d}) = P_o(\mathbf{o})\,P_d(\mathbf{d}\,|\,\mathbf{o})$ with $P_d(\mathbf{d}\,|\,\mathbf{o}) \propto P_o(\mathbf{o})$ for $\mathbf{d} \neq \mathbf{o}$. We do not allow requests with the same origin and destination, setting $P(\mathbf{d}\,|\,\mathbf{o}) = 0$ for $\mathbf{d} = \mathbf{o}$. We determine the probability $P_o(\mathbf{o})$ of an individual node $\mathbf{o}$ being the origin of a request by measuring the fraction of requests originating or terminating in the corresponding settlement over the whole duration of the pilot operation (compare Fig.~\ref{fig:FIG3_ecobus_data}b). 

We simulate the system with a fixed number of vehicles $B \in \{4,5,\ldots, 10\}$ and variable request rate $\lambda$ using the same simplified assignment and routing scheme as in our simplified model explained above. In contrast to the real-world pilot operations, we do not include rejections in our simulations. Consequently, finding shareable rides is more difficult in our simulations. On the other hand, we also neglect stop times and the spatial extension of the settlements by concentrating requests to the simplified network. This makes pooling more efficient. Despite these simplifications, we recover similar results as in our theoretical model settings and in good agreement to the empirical data (see Results below).

\clearpage

\section{Results}
We present our results in three parts. First, we introduce the \textit{load} and theoretically explain how it is the decisive quantity for estimating the reduction of distance driven of a pooled system. Second, we validate the theoretical calculation by applying it to various model simulations of ridepooling systems and demonstrate that the derived relationship holds. Third, we analyze empirical data of a ridepooling service to verify the theoretical results.

\subsection{Ridepooling load}
The dynamics of a ridepooling service fundamentally represents a queueing problem of requests entering the system and being served by the vehicle fleet. From this point of view, the \emph{load} on the system under constant external conditions quantifies the balance of the characteristic timescales of demand and supply by relating the total time $\Delta t_\mathrm{req}$ required for direct trips of all users to the total driving time $\Delta t_\mathrm{sys}$ of all vehicles in the system \citep{molkenthin2020, lotze2022}, 
\begin{equation}
    q = \frac{\Delta t_\mathrm{req}}{\Delta t_\mathrm{sys}} \,. \label{eq:load_base}
\end{equation}
In contrast to the queueing theoretical utilization factor that relates the rate at which requests enter the system with the rate with which they are served, the load $q$ is independent of the actual time required to serve a request. Instead, it directly relates demand and supply timescales with respect to motorized individual transport. The load thus quantifies the required degree of pooling to serve all requests. When $q < 1$, requests require less time to drive than overall available and could, in principle, be served by a standard ridehailing service without pooling rides. Once the load exceeds $1$, i.e.,~once requests require more time for individual service than available from the system, this is no longer possible. A single-passenger ridehailing service (and standard queueing problems) would overload. Instead, a ridepooling service starts pooling rides in the ridepooling regime $q > 1$, reducing the effective time required to serve a single request by serving multiple requests at the same time.

From this definition, it follows that the load $q$ is a lower bound for the average occupancy of the system \citep{molkenthin2020}. For example, if users request twice as much time as available from the system, the system has to serve at least two users at the same time in each vehicle. It may, however, have a higher average occupancy if users spend more time in the system due to detours or other delays during their trip. 

In terms of our toy-model, the average time requested by the system is given as the product of the average number of requests and the average trip time $\Delta t_\mathrm{trip} = \frac{\lav}{v}$ of a single request with average direct trip distance $\lav$ with constant velocity $v$. In an arbitrary reference time interval $\tau$ with an expected total number of $\lambda\tau$ requests, we thus find
\begin{equation}
    \Delta t_\mathrm{req} = \lambda\,\tau\,\Delta t_\mathrm{trip} = \lambda\,\tau\,\frac{\lav}{v}  \,. \label{eq:t_req}
\end{equation}

The total time available to the system is given by the product of the number of vehicles $B$ and the reference time interval $\tau$, 
\begin{equation}
    \Delta t_\mathrm{sys} = B\tau \,, \label{eq:tsys}
\end{equation}
Here, we only assume that all vehicles are active (or potentially available) during the whole interval, neglecting, for example, required breaks for drivers. 

The load in our simplified toy model then becomes
\begin{equation}
    q = \frac{\Delta t_\mathrm{req}}{\Delta t_\mathrm{sys}} =
    \frac{\lambda\,\tau\,\frac{\lav}{v}}{B\tau} = \frac{\lambda \lav}{vB} \,. \label{eq:load_toy_model}
\end{equation}

\newpage

\subsection{Relative distance driven}
The total distance driven, $L_\mathrm{driv}$, usually measured relative to the direct trip distance $L_\mathrm{req}$ of the requests, is a central measure to quantify the sustainability benefit of a ridepooling service. If the load $q<1$, requests are served mostly individually. Vehicles have to drive an additional distance to pick up users. In this case, ridepooling almost always results in larger distance driven compared to individual mobility by private car. However, if the load $q>1$, rides need to be pooled, thus likely reducing distance driven compared to individual mobility. We now make this relation explicit and show that the load is directly related to the relative distance driven in the ridepooling regime.

The expected total distance driven in some reference time window $\tau$ is \begin{equation}
    L_\mathrm{driv} = \left(\tau - \left<\tau_\mathrm{idle}\right>\right)\,B\,v \,, \label{eq:l_drive}
\end{equation}
where $\left<\tau_\mathrm{idle}\right>$ denotes the expected time a vehicle is idle and not driving in the period $\tau$. 
The expected total requested distance follows similar to Eq.~\eqref{eq:t_req} as
\begin{equation}
    L_\mathrm{req} = \lambda\,\tau\,\lav\,. \label{eq:l_req}
\end{equation}
Consequently, the relative distance is
\begin{equation}
    \Lrel = \frac{\left(\tau - \left<\tau_\mathrm{idle}\right>\right)\,B\,v}{\lambda\,\tau\,\lav} = \frac{1}{q} \,\left(1 - p_\mathrm{idle}\right) \leq \frac{1}{q}  \,, \label{eq:distance_saving_general}
\end{equation}
where $p_\mathrm{idle} = \left<\tau_\mathrm{idle}\right> / \tau$ denotes the probability that a vehicle of the fleet is idle. 

In the ridepooling regime, $q > 1$, where sharing trips is necessary, a ridepooling service always reduces the distance driven compared to individual mobility since $\left(1 - p_\mathrm{idle}\right) \le 1$. If the load $q$ is sufficiently large, vehicles are almost never idle, $p_\mathrm{idle} \to 0$. The load scales with relative distance as 
\begin{equation}
    \Lrel \sim \frac{1}{q} \,\label{eq:distance_saving_toy_model}
\end{equation} 
asymptotically as $q \rightarrow \infty$. In practice, this scaling approximately holds already for the load $q$ slightly above $q=1$. 
Ridepooling may also reduce the distance driven for smaller loads, depending on how many trips can be pooled without additional pickup detours, reflected in the probability of vehicles to be idle. 

\begin{figure}[ht]
    \centering
    \includegraphics{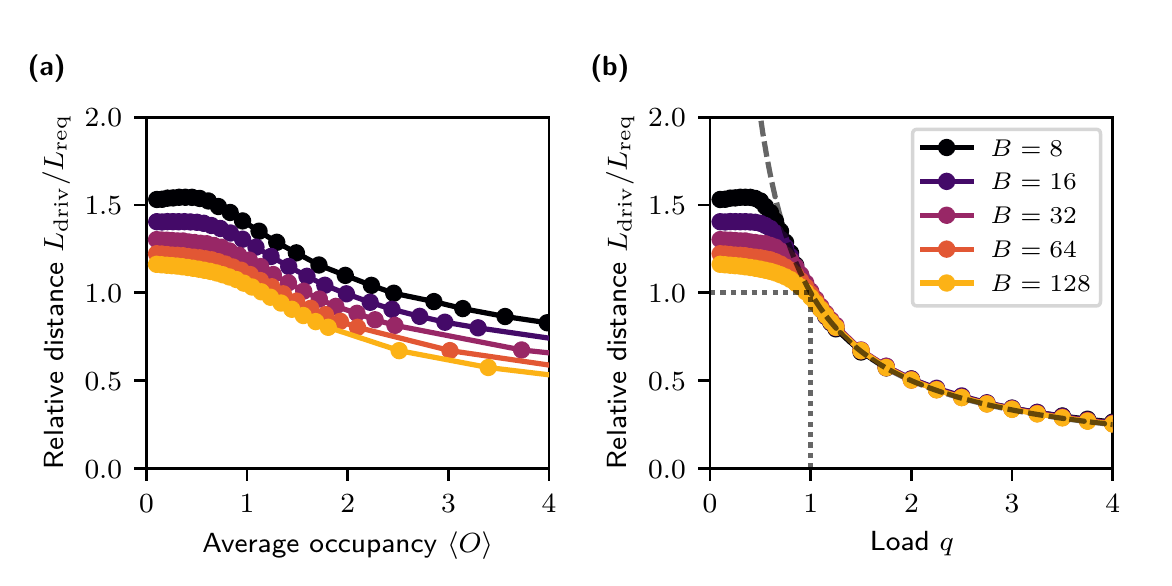}
    \caption{ 
        \textbf{Load, not occupancy, predicts relative distance driven across fleet size.}
        \textbf{a} The relative distance driven as a function of the average occupancy $\oav$ and in particular the break-even point $\Lrel = 1$ varies with the fleet size $B$ (data points show simulation results for various request rates $\lambda$). \textbf{b} In contrast, the load $q$ consistently marks the break-even point of relative distance driven, $\Lrel = 1$ at $q = 1$, for all fleet sizes (dotted lines). For all $q$, the relation $\Lrel \le \frac{1}{q}$ holds as an upper bound (dashed line).
        \vspace{-15mm}
    }
    \label{fig:FIG_fix_number_of_buses}
\end{figure}

\begin{figure}[ht]
    \centering
    \includegraphics{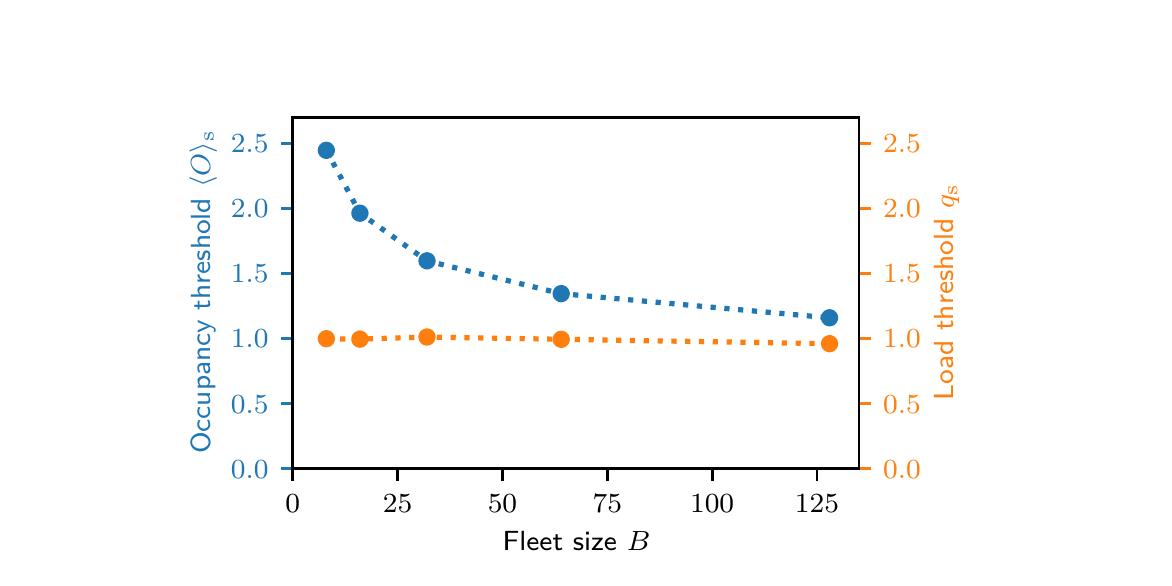}
    \caption{
        \textbf{Load consistently marks the break-even point of ridepooling sustainability.} The average occupancy $\oav_\mathrm{s}$ indicating the break-even point of distance driven compared to individual mobility (blue, computed by linear interpolation from the results shown in Fig.~\ref{fig:FIG_fix_number_of_buses}) varies with the fleet size $B$. In contrast, the load $q_\mathrm{s}$ (orange) consistently marks the break-even point at $q = 1$ independent of the fleet size. 
        \vspace{-10mm}
    }
    \label{fig:FIG_transition}
\end{figure}

\begin{figure}[ht]
    \centering
    \includegraphics{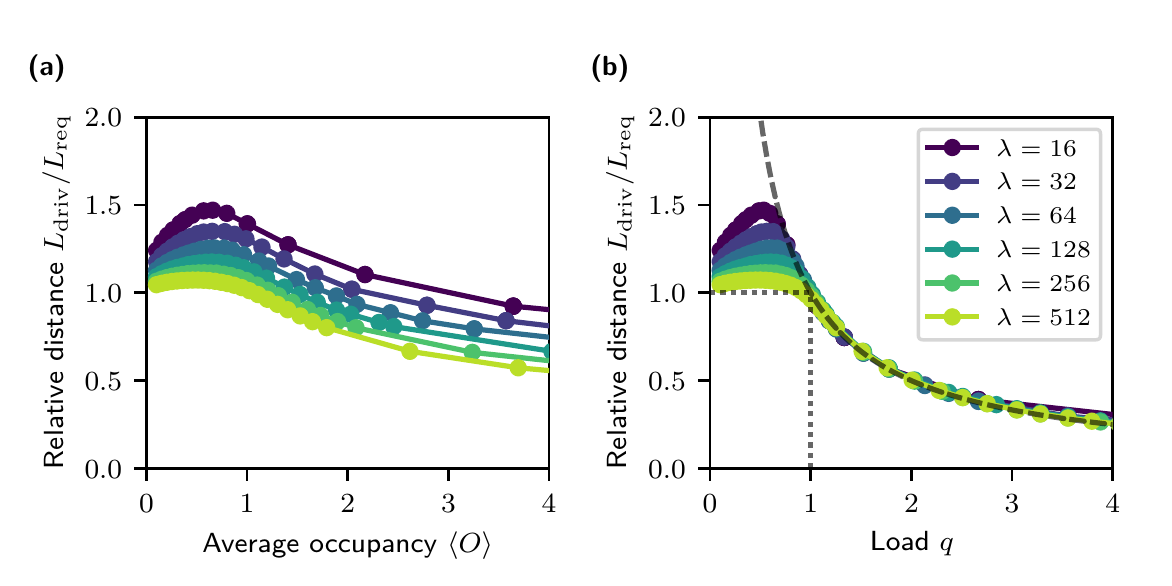}
    \caption{
        \textbf{Load, not occupancy, predicts relative distance driven across request rates.}
        \textbf{a} The relative distance driven as a function of the average occupancy $\oav$ varies with the request rate $\lambda$ (data points show simulation results for various fleet sizes $B$, compare Fig.~\ref{fig:FIG_fix_number_of_buses}). \textbf{b} In contrast, the load $q$ consistently marks the break-even point of relative distance driven, $\Lrel = 1$ at $q = 1$, for all request rates (dotted lines). For all $q$, the relation $\Lrel \le \frac{1}{q}$ holds as an upper bound (dashed line).
        \vspace{-30mm}
    }
    \label{fig:FIG_fix_request_rate}
\end{figure}

The load $q$ captures the overall collective dynamics of the ridepooling service regardless of the setting and parameters of the system, e.g., the fleet size $B$ or request rate $\lambda$, and the relation accurately holds also in finite systems \citep{molkenthin2020, lotze2022, zech2022}. In contrast to the average occupancy, the point $q=1$ directly marks the break-even point of the relative distance driven of the ridepooling service with respect to individual mobility and is easily computed from a few parameters of the ridepooling service. Figure~\ref{fig:FIG_fix_number_of_buses} and \ref{fig:FIG_transition} illustrate how the relative distance driven scales with average occupancy and load for varying request rate $\lambda$ at constant fleet size $B$. The average occupancy $\oav_\mathrm{s}$ marking the break-even point of distance driven depends explicitly on the number of buses (blue in Fig.~\ref{fig:FIG_transition}). In contrast, the break-even point in the load $q_\mathrm{s}=1$ (orange in Fig.~\ref{fig:FIG_transition}) is independent of the system setting. The same results hold when varying fleet sizes $B$ instead of request rate $\lambda$ (Fig.~\ref{fig:FIG_fix_request_rate}). 

\newpage

\newpage

\subsection{Robustness}
\begin{figure}[ht]
    \centering
    \includegraphics{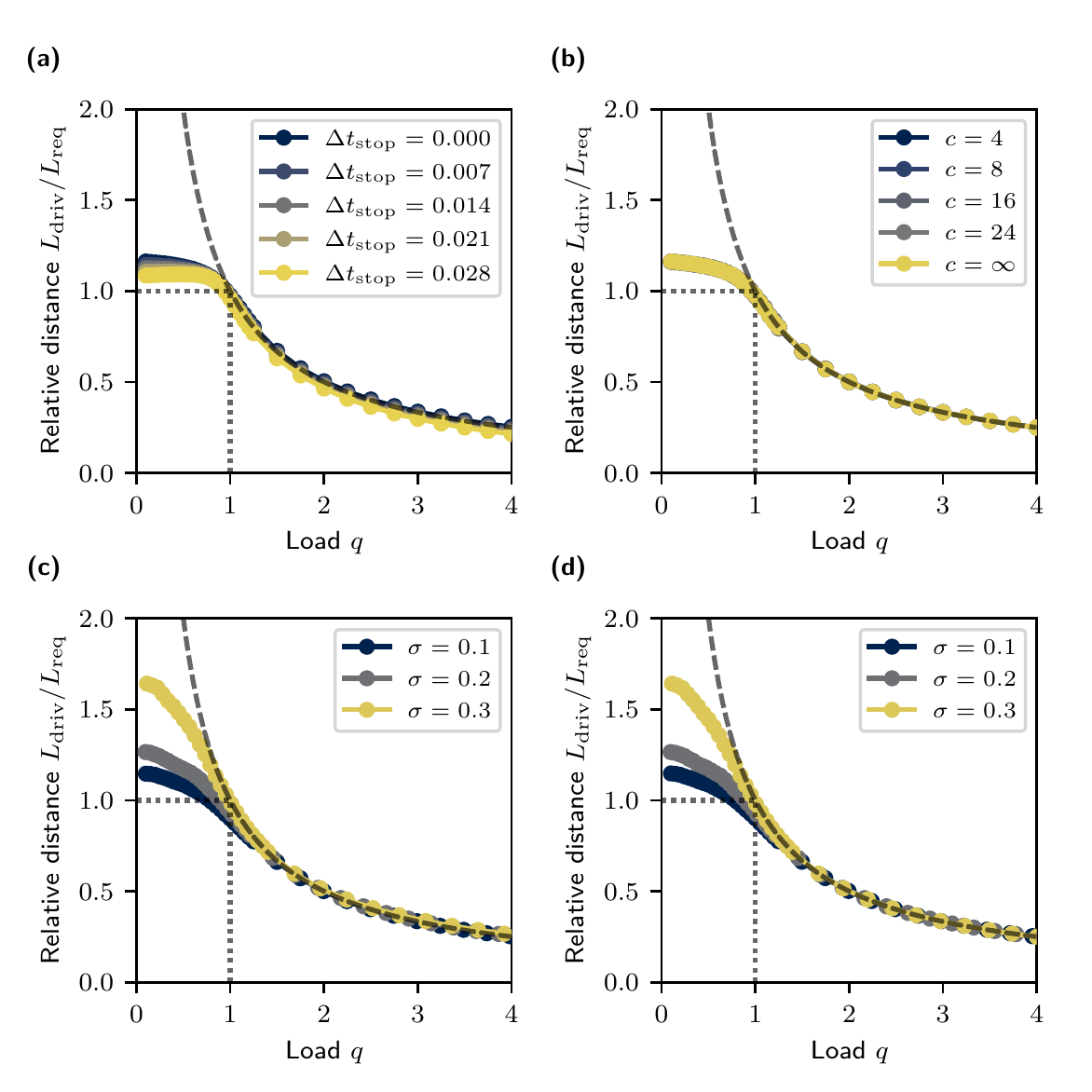}
    \caption{
        \textbf{Load robustly characterizes ridepooling sustainability.}
        The load $q$ robustly characterizes the break-even point of distance driven compared to individual mobility and predicts the relative distance driven for $q>1$ across different settings. \textbf{a} Explicit stop times (cf. Eq.~\eqref{eq:load_stoptime}). \textbf{b} Limited vehicle capacity. \textbf{c} Spatially heterogeneous demand distribution. \textbf{d} Spatially heterogeneous and asymmetric demand distribution. All panels show simulation results for a fleet size $B=128$ and varying request rate $\lambda$.
    }
    \label{fig:FIGrobustness}
\end{figure}

The concept of ridepooling load is independent of the specific ridepooling setting at hand. Here, we illustrate how the load may be extended to capture the impact of various constraints in ridepooling systems and show that the conclusions drawn from our basic model remain valid also in more complex settings (Fig.~\ref{fig:FIGrobustness}).\\

\paragraph{Stop time} 
Users require some non-zero time $\ts$ to enter and exit the vehicles at each stop. We capture this additional time in the load by starting from the definition Eq.\eqref{eq:load_base}. However, we cannot simply add this time to the requested time $\Delta t_\mathrm{req}$, since it is not shareable across multiple requests, i.e., a stop where two users exit the vehicle will take a time $2\ts$. Instead, each user entering or exiting a vehicle effectively reduces the total time available to the system for driving users to their destination. In a reference time interval $\tau$, $\tau\lambda$ requests have to enter and exit the vehicles. The available time of the system is thus reduced by $2\tau\lambda\,\ts$ to 
\begin{equation}
    \Delta t_\mathrm{sys} = B\tau-2\lambda\tau\ts = \left(B-2\lambda\,\ts\right)\,\tau\,, \label{eq:tsys_with_stop_time}
\end{equation}
Consequently, the load becomes
\begin{equation}
    q_{t_\mathrm{stop}}  = \frac{\lambda\,\lav }{v\left(B-2\lambda\,\ts\right)}\,, \label{eq:load_stoptime}
\end{equation}
which reduces to Eq.~\eqref{eq:load_toy_model} for $\ts=0$. With this adjustment, the load accurately captures the relative distance driven in terms of the relation Eq.~\eqref{eq:distance_saving_toy_model}, illustrated in Fig.~\ref{fig:FIGrobustness}a. Moreover, this extension of the load is consistent with the queuing theoretical interpretation. For large stop times $\ts \ge \frac{B}{2\lambda}$, the load diverges and the system overloads. The vehicles would need to spend all available time stopping to pick up and drop off users, and have no time to drive them to their destinations. Other non-shareable time intervals such as required breaks by drivers or refueling times may be included in the same way.\\

\paragraph{Limited capacity} 
Vehicles only have a limited capacity $c$, limiting the number of requests that can be served concurrently. This capacity limit does not change the time requested from or available to the system, and the definition of the load remains the same as in the basic model, Eq.~\eqref{eq:load_toy_model}. It does, however, affect the dynamics of the fleet and the assignment of requests to vehicles.

For loads well below the capacity $q \ll c$, the limited capacity of vehicles does not affect the dynamics significantly, as vehicles are almost never fully occupied. For loads close to but below the capacity limit, $q<c$, the system still settles into a steady state, but with a potentially very long list of waiting users and large delays, since we do not explicitly reject requests. In both cases, the load accurately describes relative distance driven of the ridepooling service in the equilibrium state (Fig.~\ref{fig:FIGrobustness}b). 
For loads above the capacity limit $q > c$, the system would overload, since it is not possible to serve sufficiently many requests at the same time. 

Overall, capacity constraints are more likely to affect the relative distance driven via external feedback mechanisms, for example reducing the (effective) request rate via rejections as the service quality reduces and delays increase when the load approaches the capacity limit. Notably, cherry-picking requests that align with the planned routes may also reduce the relative distance driven below the upper bound given by the load. \\

\paragraph{Inhomogeneous request distribution} 

Requests usually do not originate uniformly across the service area, but are more concentrated in higher density areas (e.g., city centers) and less likely in surrounding areas (suburbs or rural areas). The effect of an inhomogeneous request distribution is already captured in the load as defined in Eq.~\eqref{eq:load_toy_model} via the average trip length $\lav$. Shorter trips reduce the time requested from the system and thereby reduce the load. 

In highly heterogeneous settings with strongly concentrated demand, $\sigma = 0.1$ (see Methods and Data section for details), the relative distance driven slightly deviates from the approximate dependence predicted in Eq.~\eqref{eq:distance_saving_toy_model}. Some vehicles become idle in the low-request areas, $p_\mathrm{idle} > 0$ even for $q > 1$, reducing the total distance driven in line with Eq.~\eqref{eq:distance_saving_general}. 

Additional rebalancing of vehicles towards the center would reduce these idle times in favor of improving the service quality. Still, the ridepooling service always reduces distance driven for loads $q > 1$ (Fig.~\ref{fig:FIGrobustness}c).\\

\newpage

\paragraph{Asymmetric request distribution} 
Additionally, requests are typically not symmetric but often exhibit preferred directions that may change with the time of day. For example, more requests may go towards the city center in the morning, but more requests may leave it during the evening. As with heterogeneous, symmetric request distributions discussed in the previous paragraph, the average trip length $\lav$ captures the effects of the demand distribution. While the occupancy of the vehicles differs for trips in different directions (e.g., serving two requests in one direction and returning empty in the other for an average occupancy of one), the relative distance driven is still captured by the load as defined in Eq.~\eqref{eq:load_toy_model}, illustrated in Fig.~\ref{fig:FIGrobustness}d. Small deviations for highly heterogeneous settings occur due to idle vehicles in low request areas, as in the previous example.

\clearpage

\subsection{Evaluating demand-responsive ridepooling on empirical data}
Evaluating empirical data from a pilot operation of an on-demand ridepooling service that served as an alternative to infrequent line-based public transport in a rural area of Germany (see Methods and Data for details) further validates our results. We compute all quantities of interest by accumulating data for every hour that the service was active and taking the arithmetic mean: the total distance driven $L_\mathrm{driv}$, the total requested distance $L_\mathrm{req}$, the load $q$, and the average occupancy of the vehicles $\oav$. The empirical request rate used to compute the load is defined as the number of pick-ups that occur during each hour, approximating the rate of served requests.

\begin{figure}[ht]
    \includegraphics{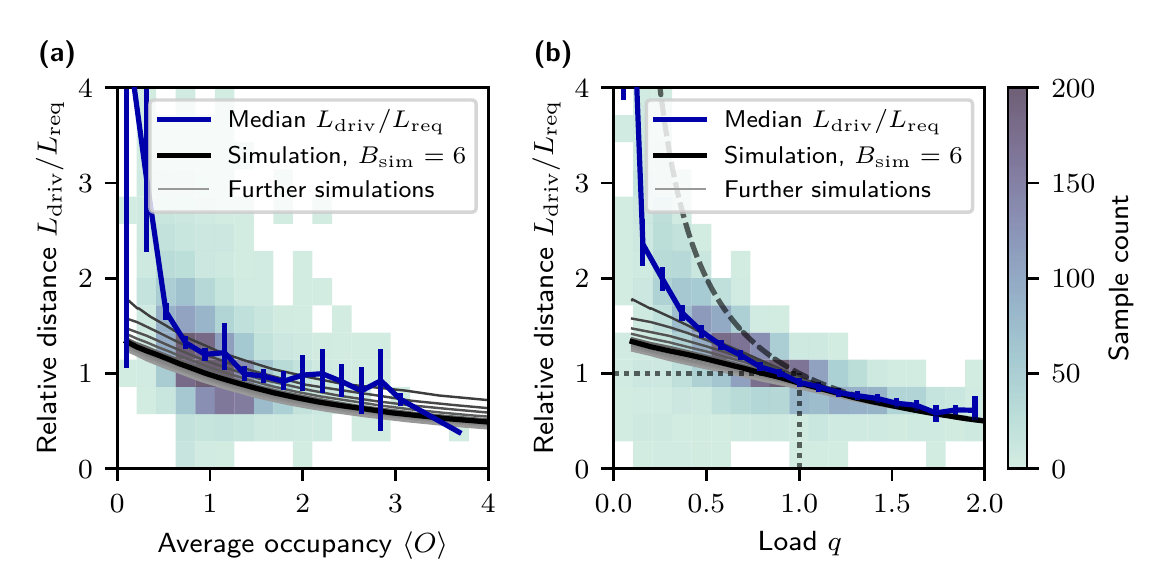}
    \caption{
    \textbf{Load, not occupancy, predicts relative distance driven for empirical ridepooling data.} 
    \textbf{a} Hourly averages of the relative distance driven across the service period vary strongly, even for similar average occupancy $\oav$ (2D histogram, blue squares). The occupancy is normalized with the average number of passengers per request, since the empirical requests may include multiple passengers. \textbf{b} In contrast, variations of the relative distance driven as a function of the load $q$ are smaller and the relative distance driven is consistently bounded by $\Lrel \le 1/q$ (dashed line) for the majority of the observations. Aggregated median values of the relative distance driven with $95\,\%$ confidence intervals computed by bootstrapping (blue line) are roughly approximated by model simulations with different fleet sizes (gray, typical fleet size $B=6$ of the pilot operation in black, see Methods and Data for details). 
    \vspace{-15mm}
    }
    \label{fig:FIGZ_ecobus_load}
\end{figure}

Figure \ref{fig:FIGZ_ecobus_load} shows the distribution of observations of the relative distance driven $\Lrel$ against both the average occupancy $\oav$ and the load $q$. Observations of the relative distance as a function of the average occupancy $\oav$ are broadly distributed (Fig.~\ref{fig:FIGZ_ecobus_load}a), similar to our theoretical results above and the model simulations (gray lines representing simulation results with different fleet sizes, thick black line representing the typical fleet size $B = 6$). While the service generally reduces distance driven for larger occupancy values, no clear transition to $\Lrel < 1$ is visible. As a function of the load $q$, however, fluctuations in the relative distance driven become smaller and the data collapses more strongly on the expected curve Eq.~\eqref{eq:distance_saving_toy_model} as in the model simulations (Fig.~\ref{fig:FIGZ_ecobus_load}b), reducing the distance driven for $q > 1$. For small loads, partially idling vehicles reduce the expected distance driven compared to the upper bound given by the load, as also discussed in the previous model extensions. 

\begin{figure}[ht]
    \includegraphics{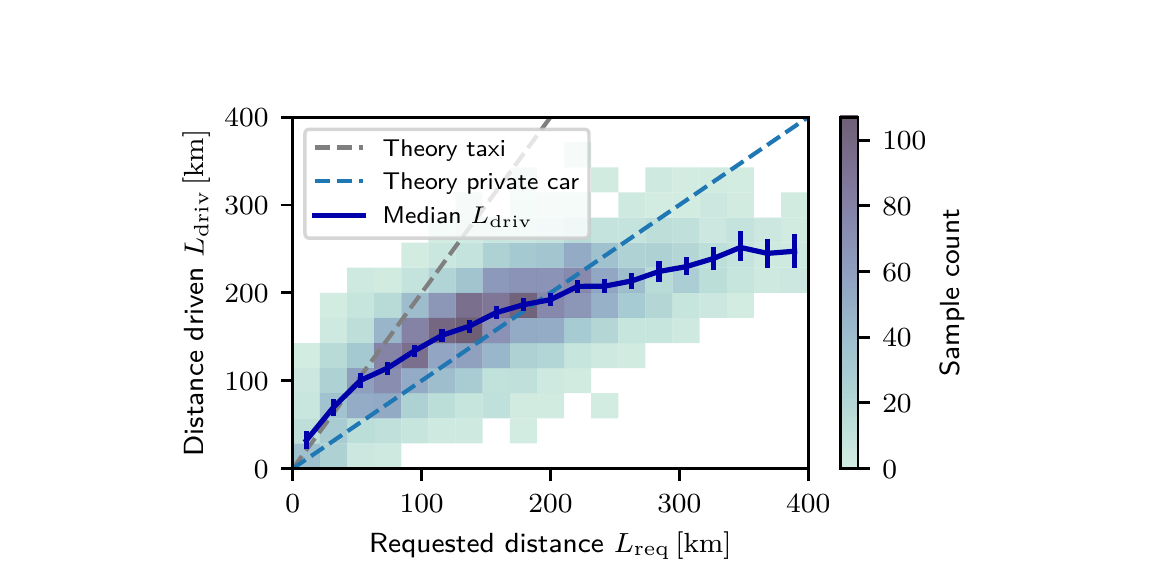}
    \caption{
        \textbf{Transition to ridepooling sustainability.}
        For small requested distances (few requests), ridepooling behaves like a single-passenger ridehailing service (taxi, blue dashed line), increasing the distance driven due to empty trips to pick up users. For large requested distances (many requests), ridepooling reduces the distance driven. Even in the small-scale pilot operation ($B \le 10$ vehicles) ridepooling reduced the total distance driven compared to individual mobility (gray dashed line) during times of high demand, above $L_\mathrm{req} \approx 200\,\mathrm{km}$. The blue line shows the aggregated median of the hourly distance driven with $95\,\%$ confidence intervals computed by bootstrapping the data, the 2D histogram in the background shows the full dataset (compare Fig.~\ref{fig:FIGZ_ecobus_load}).
        }
    \label{fig:FIGZ_ecobus_distance}
\end{figure}

The relative distance driven observed in the empirical data varies strongly (Fig.~\ref{fig:FIGZ_ecobus_distance}). For few requests per hour (small requested distance), the service required approximately twice the requested distance. During these times, the service operated as an inefficient ridehailing service with equal amounts of passenger trips and deadheading to pick up passengers. However, during times of high request rate (large requested distance), the service reduced the distance driven by up to $40\,\%$ compared to the requested distance. 

Variations around the expected scaling stem from various complications of evaluating the empirical data. Since conditions are not stationary as in our model simulations, the binning of the data results in mismatches between the distance driven and the request rate in one hour. A request in one hourly bin may contribute both to the distance driven during that hour, but also to the distance driven during the next hour. The variable request rates thus result in deviations compared to our idealized stationary model simulations. Additionally, variation of the fleet size and the service conditions such as the pre-booking of requests during the operation period and the large number of rejections (compare the description of the service in Methods and Data) further increase the variability.

Overall, the load $q$ serves a good predictor and upper bound of the relative distance driven also for the empirical data. Importantly, the expected load of the service could be easily computed from the served request rate, average distance, typical velocity, and the fleet size. Detailed simulations or direct empirical observations are not required, highlighting its usefulness as a predictor for ridepooling sustainability.
\vspace{-5mm}

\section{Conclusion}
The relative distance driven $\Lrel$ is a key observable of a ridepooling operation relating the total distance that all service vehicles drive along their routes to the sum of the direct distances of all requested trips, i.e., the total distance driven if all trips were driven individually by private car. The relative distance thus quantifies the ecological sustainability of pooled systems: At $\Lrel < 1$, ridepooling is more sustainable than individual transport in terms of total distance driven, at $\Lrel > 1$ it is less sustainable. While the relative distance driven is related to other key performance indicators like the average vehicle occupancy, each indicator captures a specific aspect of the operational efficiency of a ridepooling service. Other key performance indicators thus cannot be directly interpreted in terms of the ecological sustainability of the service.

We have presented the ridepooling load $q$, a system-level dimensionless parameter quantifying the ratio of characteristic demand and supply timescales. The load provides an upper bound for the relative distance driven, $\Lrel \le 1/q$, Eq.~\eqref{eq:distance_saving_toy_model}, such that $q = 1$ indicates the break-even point above which ridepooling vehicles drive less distance in total than passengers would with private cars. This result is independent of other details of the service or the setting, such as capacity limits or the spatial request distribution. Additionally, the upper bound $1/q$ provides an accurate estimate of the relative distance in the ridepooling regime, $q > 1$, where trips need to be shared to serve all requests. Detailed event-based simulations of a model ridepooling system in various settings and data-driven analysis of a real-world ridepooling operation demonstrate the robustness of these relations.
 
Importantly, the load $q$ depends only on aggregate properties that may be easily estimated as a function of time of day, day of week, service area, expected or existing demand and other settings. Specifically, it combines the fleet size, the request rate of trips, their average distances, and a typical vehicle velocity. In contrast to standard key performance indicators, this permits straightforward evaluation of the load even without direct simulation, operation, or observation of a ridepooling service. The load thus enables service providers and policymakers to quickly gauge whether a ridepooling operation in a given setting is likely to reduce the total distance driven.

\newpage

\section{Discussion}
As defined, $q$ quantifies the load of a ridehailing or ridepooling service, similar to the load factor of a queuing system, relative to individual mobility or single-passenger ridehailing. It thus naturally compares ridepooling to direct trips, explaining the connection to the relative distance driven. Following the idea of the load as a systemic indicator, several additional effects in more complex settings are already captured or may be easily included, as illustrated by our model extensions with stop times, limited vehicle capacity, or non-uniform request distributions, compare also \citep{molkenthin2020, alonso-mora2017, zech2022}. 

However, the load alone is not sufficient to fully characterize the dynamics of the ridepooling service. Waiting times or delays of users also crucially depend on the scale of the service, i.e., the total number of requests and vehicles. Larger, busier systems typically offer more opportunities to pool requests with little detour. Other details, such as the structure of the underlying street network or the demand distribution, that also affect these quantities are already approximately captured by the load in terms of the request rate and the average direct trip distance. Similar definitions of effective loads and other characteristic quantities may enable more explicit connections to other observables or quantities of interest, or to describe the dynamics relative to other reference states \citep{molkenthin2020, tachet2017}. 

The load of a ridepooling system quantifies the steady state dynamics of the vehicle fleet. It relies on well-defined average values for the number of requests, the drive time, and the typical vehicle velocity. The load thus captures the ridepooling dynamics on sufficiently long timescales when service conditions are roughly constant over timescales needed to serve one passenger request. If the request rate or other system settings change on faster scales, such as during a fast increase of the number of requests during morning rush-hour, the dynamics of the vehicle fleet will not reflect the equilibrium state predicted by the (instantaneous) load. This effect is partly responsible for the spread of the empirical observations (Fig.~\ref{fig:FIGZ_ecobus_load}): when many requests are made in the later part of an hourly interval (high request rate and high load) but are only served in the next interval (low request rate and low load), the instantaneous quantification of the load does not reflect the state of the service in the second interval. Similar deviations from the theoretical prediction may occur when the request distribution or other parameters of the service change more quickly than the intrinsic equilibration timescale of the service, for example due to strongly varying vehicle velocities in congested traffic. Consequently, the exact mathematical inequality derived above, Eq.~\eqref{eq:distance_saving_toy_model}, may not hold for small timescales. The load still serves as an indicator for the achievable sustainability of ridepooling in typical service conditions.

Overall, our results demonstrate how measures based on the fundamental collective dynamics on the systems' level and in particular dimensionless parameters, such as the ridepooling load introduced above, may complement detailed case studies of specific individual scenarios and contribute to a better understanding of ridepooling services and comparability of different settings and service conditions.

\newpage

\subsection*{Acknowledgments}
The authors thank Verena Krall and all members of the Chair of Network Dynamics for valuable discussions. CL acknowledges support from the German Federal Environmental Foundation (Deutsche Bundesstiftung Umwelt DBU). This research was partially supported by the German Federal Ministry for Education and Research (BMBF) under grant number 16ICR01, 
by the Südniedersachsenprogramm of the of State of Lower Saxony and the European Community within the European Fund for Regional Development (EFRE) under grant number 85003731. 
This research was supported through the Center for Advancing Electronics Dresden (cfaed). The authors are grateful to the Center for Information Services and High Performance Computing (ZIH) TU Dresden for providing facilities for high throughput calculations.

\subsection*{Competing Interests}
DM was employed at MOIA GmbH when the research was conducted. MOIA GmbH neither sponsored nor endorses his research.

\subsection*{Author contributions}
CL: Conceptualization, literature review, simulations and data analysis, manuscript writing, and editing. 
PM: Simulations and data analysis, literature review, manuscript writing, and editing. 
FJ: Simulations and data analysis, literature review, manuscript writing, and editing. 
DM: data analysis and data interpretation, manuscript editing
MT: Conceptualization, data interpretation, manuscript writing, and editing. 
MS: Conceptualization, modeling, data interpretation, manuscript writing, and editing. 

\subsection*{Code and data availability}
Data and code are available from the authors. The original data from the EcoBus pilot operation are not publicly available.


\begin{thebibliography}{41}%
\makeatletter
\providecommand \@ifxundefined [1]{%
 \@ifx{#1\undefined}
}%
\providecommand \@ifnum [1]{%
 \ifnum #1\expandafter \@firstoftwo
 \else \expandafter \@secondoftwo
 \fi
}%
\providecommand \@ifx [1]{%
 \ifx #1\expandafter \@firstoftwo
 \else \expandafter \@secondoftwo
 \fi
}%
\providecommand \natexlab [1]{#1}%
\providecommand \enquote  [1]{``#1''}%
\providecommand \bibnamefont  [1]{#1}%
\providecommand \bibfnamefont [1]{#1}%
\providecommand \citenamefont [1]{#1}%
\providecommand \href@noop [0]{\@secondoftwo}%
\providecommand \href [0]{\begingroup \@sanitize@url \@href}%
\providecommand \@href[1]{\@@startlink{#1}\@@href}%
\providecommand \@@href[1]{\endgroup#1\@@endlink}%
\providecommand \@sanitize@url [0]{\catcode `\\12\catcode `\$12\catcode
  `\&12\catcode `\#12\catcode `\^12\catcode `\_12\catcode `\%12\relax}%
\providecommand \@@startlink[1]{}%
\providecommand \@@endlink[0]{}%
\providecommand \url  [0]{\begingroup\@sanitize@url \@url }%
\providecommand \@url [1]{\endgroup\@href {#1}{\urlprefix }}%
\providecommand \urlprefix  [0]{URL }%
\providecommand \Eprint [0]{\href }%
\providecommand \doibase [0]{https://doi.org/}%
\providecommand \selectlanguage [0]{\@gobble}%
\providecommand \bibinfo  [0]{\@secondoftwo}%
\providecommand \bibfield  [0]{\@secondoftwo}%
\providecommand \translation [1]{[#1]}%
\providecommand \BibitemOpen [0]{}%
\providecommand \bibitemStop [0]{}%
\providecommand \bibitemNoStop [0]{.\EOS\space}%
\providecommand \EOS [0]{\spacefactor3000\relax}%
\providecommand \BibitemShut  [1]{\csname bibitem#1\endcsname}%
\let\auto@bib@innerbib\@empty
\bibitem [{\citenamefont {Santi}\ \emph {et~al.}(2014)\citenamefont {Santi},
  \citenamefont {Resta}, \citenamefont {Szell}, \citenamefont {Sobolevsky},
  \citenamefont {Strogatz},\ and\ \citenamefont {Ratti}}]{santi2014}%
  \BibitemOpen
  \bibfield  {author} {\bibinfo {author} {\bibfnamefont {P.}~\bibnamefont
  {Santi}}, \bibinfo {author} {\bibfnamefont {G.}~\bibnamefont {Resta}},
  \bibinfo {author} {\bibfnamefont {M.}~\bibnamefont {Szell}}, \bibinfo
  {author} {\bibfnamefont {S.}~\bibnamefont {Sobolevsky}}, \bibinfo {author}
  {\bibfnamefont {S.~H.}\ \bibnamefont {Strogatz}},\ and\ \bibinfo {author}
  {\bibfnamefont {C.}~\bibnamefont {Ratti}},\ }\bibfield  {title} {\bibinfo
  {title} {Quantifying the {{Benefits}} of {{Vehicle Pooling}} with
  {{Shareability Networks}}},\ }\href {https://doi.org/10.1073/pnas.1403657111}
  {\bibfield  {journal} {\bibinfo  {journal} {Proc. Natl. Acad. Sci.}\ }\textbf
  {\bibinfo {volume} {111}},\ \bibinfo {pages} {13290} (\bibinfo {year}
  {2014})}\BibitemShut {NoStop}%
\bibitem [{\citenamefont {{Alonso-Mora}}\ \emph {et~al.}(2017)\citenamefont
  {{Alonso-Mora}}, \citenamefont {Samaranayake}, \citenamefont {Wallar},
  \citenamefont {Frazzoli},\ and\ \citenamefont {Rus}}]{alonso-mora2017}%
  \BibitemOpen
  \bibfield  {author} {\bibinfo {author} {\bibfnamefont {J.}~\bibnamefont
  {{Alonso-Mora}}}, \bibinfo {author} {\bibfnamefont {S.}~\bibnamefont
  {Samaranayake}}, \bibinfo {author} {\bibfnamefont {A.}~\bibnamefont
  {Wallar}}, \bibinfo {author} {\bibfnamefont {E.}~\bibnamefont {Frazzoli}},\
  and\ \bibinfo {author} {\bibfnamefont {D.}~\bibnamefont {Rus}},\ }\bibfield
  {title} {\bibinfo {title} {On-{{Demand High-Capacity Ride-Sharing}} via
  {{Dynamic Trip-Vehicle Assignment}}},\ }\href
  {https://doi.org/10.1073/pnas.1611675114} {\bibfield  {journal} {\bibinfo
  {journal} {Proc. Natl. Acad. Sci.}\ }\textbf {\bibinfo {volume} {114}},\
  \bibinfo {pages} {462} (\bibinfo {year} {2017})}\BibitemShut {NoStop}%
\bibitem [{\citenamefont {Tachet}\ \emph {et~al.}(2017)\citenamefont {Tachet},
  \citenamefont {Sagarra}, \citenamefont {Santi}, \citenamefont {Resta},
  \citenamefont {Szell}, \citenamefont {Strogatz},\ and\ \citenamefont
  {Ratti}}]{tachet2017}%
  \BibitemOpen
  \bibfield  {author} {\bibinfo {author} {\bibfnamefont {R.}~\bibnamefont
  {Tachet}}, \bibinfo {author} {\bibfnamefont {O.}~\bibnamefont {Sagarra}},
  \bibinfo {author} {\bibfnamefont {P.}~\bibnamefont {Santi}}, \bibinfo
  {author} {\bibfnamefont {G.}~\bibnamefont {Resta}}, \bibinfo {author}
  {\bibfnamefont {M.}~\bibnamefont {Szell}}, \bibinfo {author} {\bibfnamefont
  {S.~H.}\ \bibnamefont {Strogatz}},\ and\ \bibinfo {author} {\bibfnamefont
  {C.}~\bibnamefont {Ratti}},\ }\bibfield  {title} {\bibinfo {title} {Scaling
  {{Law}} of {{Urban Ride Sharing}}},\ }\href
  {https://doi.org/10.1038/srep42868} {\bibfield  {journal} {\bibinfo
  {journal} {Sci. Rep.}\ }\textbf {\bibinfo {volume} {7}},\ \bibinfo {pages}
  {42868} (\bibinfo {year} {2017})}\BibitemShut {NoStop}%
\bibitem [{\citenamefont {Zwick}\ \emph
  {et~al.}(2021{\natexlab{a}})\citenamefont {Zwick}, \citenamefont {Kuehnel},
  \citenamefont {Moeckel},\ and\ \citenamefont {Axhausen}}]{zwick2021ride}%
  \BibitemOpen
  \bibfield  {author} {\bibinfo {author} {\bibfnamefont {F.}~\bibnamefont
  {Zwick}}, \bibinfo {author} {\bibfnamefont {N.}~\bibnamefont {Kuehnel}},
  \bibinfo {author} {\bibfnamefont {R.}~\bibnamefont {Moeckel}},\ and\ \bibinfo
  {author} {\bibfnamefont {K.~W.}\ \bibnamefont {Axhausen}},\ }\bibfield
  {title} {\bibinfo {title} {Ride-pooling efficiency in large, medium-sized and
  small towns-simulation assessment in the munich metropolitan region},\ }\href
  {https://doi.org/10.1016/j.procs.2021.03.083} {\bibfield  {journal} {\bibinfo
   {journal} {Procedia Comput. Sci.}\ }\textbf {\bibinfo {volume} {184}},\
  \bibinfo {pages} {662} (\bibinfo {year} {2021}{\natexlab{a}})}\BibitemShut
  {NoStop}%
\bibitem [{\citenamefont {Storch}\ \emph {et~al.}(2021)\citenamefont {Storch},
  \citenamefont {Timme},\ and\ \citenamefont {Schr{\"o}der}}]{storch2021}%
  \BibitemOpen
  \bibfield  {author} {\bibinfo {author} {\bibfnamefont {D.-M.}\ \bibnamefont
  {Storch}}, \bibinfo {author} {\bibfnamefont {M.}~\bibnamefont {Timme}},\ and\
  \bibinfo {author} {\bibfnamefont {M.}~\bibnamefont {Schr{\"o}der}},\
  }\bibfield  {title} {\bibinfo {title} {Incentive-driven transition to high
  ride-sharing adoption},\ }\href {https://doi.org/10.1038/s41467-021-23287-6}
  {\bibfield  {journal} {\bibinfo  {journal} {Nat. Commun.}\ }\textbf {\bibinfo
  {volume} {12}},\ \bibinfo {pages} {3003} (\bibinfo {year}
  {2021})}\BibitemShut {NoStop}%
\bibitem [{\citenamefont {EcoBus}(2019)}]{ecobus}%
  \BibitemOpen
  \bibfield  {author} {\bibinfo {author} {\bibnamefont {EcoBus}},\ }\href
  {https://projekt.ecobus-online.de/home.html} {\bibinfo {title} {Ecobus
  ridepooling pilot project}} (\bibinfo {year} {2019}),\ \bibinfo {note}
  {\url{https://projekt.ecobus-online.de/home.html}}\BibitemShut {NoStop}%
\bibitem [{\citenamefont {B{\"u}rstlein}\ \emph {et~al.}(2021)\citenamefont
  {B{\"u}rstlein}, \citenamefont {L{\'o}pez},\ and\ \citenamefont
  {Farooq}}]{burstlein2021}%
  \BibitemOpen
  \bibfield  {author} {\bibinfo {author} {\bibfnamefont {J.}~\bibnamefont
  {B{\"u}rstlein}}, \bibinfo {author} {\bibfnamefont {D.}~\bibnamefont
  {L{\'o}pez}},\ and\ \bibinfo {author} {\bibfnamefont {B.}~\bibnamefont
  {Farooq}},\ }\bibfield  {title} {\bibinfo {title} {Exploring {{First-Mile}}
  on-{{Demand Transit Solutions}} for {{North American Suburbia}}: {{A Case
  Study}} of {{Markham}}, {{Canada}}},\ }\href@noop {} {\bibfield  {journal}
  {\bibinfo  {journal} {Transp. Res. A}\ }\textbf {\bibinfo {volume} {153}},\
  \bibinfo {pages} {261} (\bibinfo {year} {2021})}\BibitemShut {NoStop}%
\bibitem [{\citenamefont {Kaddoura}\ \emph {et~al.}(2021)\citenamefont
  {Kaddoura}, \citenamefont {Leich}, \citenamefont {Neumann},\ and\
  \citenamefont {Nagel}}]{kaddoura2021}%
  \BibitemOpen
  \bibfield  {author} {\bibinfo {author} {\bibfnamefont {I.}~\bibnamefont
  {Kaddoura}}, \bibinfo {author} {\bibfnamefont {G.}~\bibnamefont {Leich}},
  \bibinfo {author} {\bibfnamefont {A.}~\bibnamefont {Neumann}},\ and\ \bibinfo
  {author} {\bibfnamefont {K.}~\bibnamefont {Nagel}},\ }\bibfield  {title}
  {\bibinfo {title} {From today's ride-sharing services to future mobility
  concepts: {{A}} simulation study for urban and rural areas},\ }\bibfield
  {journal} {\bibinfo  {journal} {Preprint}\ }\href
  {https://doi.org/10.14279/depositonce-12055} {10.14279/depositonce-12055}
  (\bibinfo {year} {2021})\BibitemShut {NoStop}%
\bibitem [{\citenamefont {Erhardt}\ \emph {et~al.}(2019)\citenamefont
  {Erhardt}, \citenamefont {Roy}, \citenamefont {Cooper}, \citenamefont {Sana},
  \citenamefont {Chen},\ and\ \citenamefont {Castiglione}}]{erhardt2019}%
  \BibitemOpen
  \bibfield  {author} {\bibinfo {author} {\bibfnamefont {G.~D.}\ \bibnamefont
  {Erhardt}}, \bibinfo {author} {\bibfnamefont {S.}~\bibnamefont {Roy}},
  \bibinfo {author} {\bibfnamefont {D.}~\bibnamefont {Cooper}}, \bibinfo
  {author} {\bibfnamefont {B.}~\bibnamefont {Sana}}, \bibinfo {author}
  {\bibfnamefont {M.}~\bibnamefont {Chen}},\ and\ \bibinfo {author}
  {\bibfnamefont {J.}~\bibnamefont {Castiglione}},\ }\bibfield  {title}
  {\bibinfo {title} {Do {{Transportation Network Companies Decrease}} or
  {{Increase Congestion}}?},\ }\href {https://doi.org/10.1126/sciadv.aau2670}
  {\bibfield  {journal} {\bibinfo  {journal} {Sci. Adv.}\ }\textbf {\bibinfo
  {volume} {5}},\ \bibinfo {pages} {eaau2670} (\bibinfo {year}
  {2019})}\BibitemShut {NoStop}%
\bibitem [{\citenamefont {Henao}\ and\ \citenamefont
  {Marshall}(2019)}]{henao2019}%
  \BibitemOpen
  \bibfield  {author} {\bibinfo {author} {\bibfnamefont {A.}~\bibnamefont
  {Henao}}\ and\ \bibinfo {author} {\bibfnamefont {W.~E.}\ \bibnamefont
  {Marshall}},\ }\bibfield  {title} {\bibinfo {title} {The {{Impact}} of
  {{Ride-Hailing}} on {{Vehicle Miles Traveled}}},\ }\href
  {https://doi.org/10.1007/s11116-018-9923-2} {\bibfield  {journal} {\bibinfo
  {journal} {Transportation}\ }\textbf {\bibinfo {volume} {46}},\ \bibinfo
  {pages} {2173} (\bibinfo {year} {2019})}\BibitemShut {NoStop}%
\bibitem [{\citenamefont {Erhardt}\ \emph {et~al.}(2021)\citenamefont
  {Erhardt}, \citenamefont {Mucci}, \citenamefont {Cooper}, \citenamefont
  {Sana}, \citenamefont {Chen},\ and\ \citenamefont
  {Castiglione}}]{erhardt2021}%
  \BibitemOpen
  \bibfield  {author} {\bibinfo {author} {\bibfnamefont {G.~D.}\ \bibnamefont
  {Erhardt}}, \bibinfo {author} {\bibfnamefont {R.~A.}\ \bibnamefont {Mucci}},
  \bibinfo {author} {\bibfnamefont {D.}~\bibnamefont {Cooper}}, \bibinfo
  {author} {\bibfnamefont {B.}~\bibnamefont {Sana}}, \bibinfo {author}
  {\bibfnamefont {M.}~\bibnamefont {Chen}},\ and\ \bibinfo {author}
  {\bibfnamefont {J.}~\bibnamefont {Castiglione}},\ }\bibfield  {title}
  {\bibinfo {title} {Do {{Transportation Network Companies Increase}} or
  {{Decrease Transit Ridership}}? {{Empirical Evidence}} from {{San
  Francisco}}},\ }\href {https://doi.org/10.1007/s11116-021-10178-4} {\bibfield
   {journal} {\bibinfo  {journal} {Transportation}\ }\textbf {\bibinfo {volume}
  {49}},\ \bibinfo {pages} {313–342} (\bibinfo {year} {2021})}\BibitemShut
  {NoStop}%
\bibitem [{\citenamefont {Diao}\ \emph {et~al.}(2021)\citenamefont {Diao},
  \citenamefont {Kong},\ and\ \citenamefont {Zhao}}]{diao2021}%
  \BibitemOpen
  \bibfield  {author} {\bibinfo {author} {\bibfnamefont {M.}~\bibnamefont
  {Diao}}, \bibinfo {author} {\bibfnamefont {H.}~\bibnamefont {Kong}},\ and\
  \bibinfo {author} {\bibfnamefont {J.}~\bibnamefont {Zhao}},\ }\bibfield
  {title} {\bibinfo {title} {Impacts of {{Transportation Network Companies}} on
  {{Urban Mobility}}},\ }\href {https://doi.org/10.1038/s41893-020-00678-z}
  {\bibfield  {journal} {\bibinfo  {journal} {Nat. Sustain.}\ }\textbf
  {\bibinfo {volume} {4}},\ \bibinfo {pages} {494–500} (\bibinfo {year}
  {2021})}\BibitemShut {NoStop}%
\bibitem [{\citenamefont {Liebchen}\ \emph {et~al.}(2021)\citenamefont
  {Liebchen}, \citenamefont {Lehnert}, \citenamefont {Mehlert},\ and\
  \citenamefont {Schiefelbusch}}]{liebchen2021}%
  \BibitemOpen
  \bibfield  {author} {\bibinfo {author} {\bibfnamefont {C.}~\bibnamefont
  {Liebchen}}, \bibinfo {author} {\bibfnamefont {M.}~\bibnamefont {Lehnert}},
  \bibinfo {author} {\bibfnamefont {C.}~\bibnamefont {Mehlert}},\ and\ \bibinfo
  {author} {\bibfnamefont {M.}~\bibnamefont {Schiefelbusch}},\ }\bibfield
  {title} {\bibinfo {title} {{Ridepooling-Effizienz messbar machen}},\ }in\
  \href {http://link.springer.com/10.1007/978-3-658-32266-3\_7} {\emph
  {\bibinfo {booktitle} {{Making Connected Mobility Work}}}},\ \bibinfo
  {editor} {edited by\ \bibinfo {editor} {\bibfnamefont {H.}~\bibnamefont
  {Proff}}}\ (\bibinfo  {publisher} {{Springer Fachmedien Wiesbaden}},\
  \bibinfo {address} {{Wiesbaden}},\ \bibinfo {year} {2021})\ pp.\ \bibinfo
  {pages} {135--150}\BibitemShut {NoStop}%
\bibitem [{\citenamefont {Wolfe}\ and\ \citenamefont
  {Levine}(2018)}]{wolfe2018}%
  \BibitemOpen
  \bibfield  {author} {\bibinfo {author} {\bibfnamefont {J.}~\bibnamefont
  {Wolfe}}\ and\ \bibinfo {author} {\bibfnamefont {A.~S.}\ \bibnamefont
  {Levine}},\ }\bibfield  {title} {\bibinfo {title} {New {{York Today}}:
  {{Capping Uber}}},\ }\href
  {https://www.nytimes.com/2018/08/15/nyregion/new-york-today-
  sunglasses-eye-safety.html} {\bibfield  {journal} {\bibinfo  {journal} {The
  New York Times}\ } (\bibinfo {year} {2018})}\BibitemShut {NoStop}%
\bibitem [{\citenamefont {Honan}(2019)}]{honan2019}%
  \BibitemOpen
  \bibfield  {author} {\bibinfo {author} {\bibfnamefont {K.}~\bibnamefont
  {Honan}},\ }\bibfield  {title} {\bibinfo {title} {Uber, {{Lyft Drivers Face
  Stiffer Regulations}} in {{New York City}}},\ }\href
  {https://www.wsj.com/articles/uber-lyft-drivers-face-stiffer-
  regulations-in-new-york-city-11560375449} {\bibfield  {journal} {\bibinfo
  {journal} {Wall Street Journal}\ } (\bibinfo {year} {2019})}\BibitemShut
  {NoStop}%
\bibitem [{\citenamefont {{PBefG}}(2021)}]{pbefg2021}%
  \BibitemOpen
  \bibfield  {author} {\bibinfo {author} {\bibnamefont {{PBefG}}},\ }\href
  {https://www.gesetze-im-internet.de/pbefg/\_\_50.html} {\bibinfo {title}
  {Personenbef\"orderungsgesetz ({{PBefG}}) - \textsection{} 50 {{Geb\"undelter
  Bedarfsverkehr}}}} (\bibinfo {year} {2021})\BibitemShut {NoStop}%
\bibitem [{\citenamefont {Molkenthin}\ \emph {et~al.}(2020)\citenamefont
  {Molkenthin}, \citenamefont {Schr{\"o}der},\ and\ \citenamefont
  {Timme}}]{molkenthin2020}%
  \BibitemOpen
  \bibfield  {author} {\bibinfo {author} {\bibfnamefont {N.}~\bibnamefont
  {Molkenthin}}, \bibinfo {author} {\bibfnamefont {M.}~\bibnamefont
  {Schr{\"o}der}},\ and\ \bibinfo {author} {\bibfnamefont {M.}~\bibnamefont
  {Timme}},\ }\bibfield  {title} {\bibinfo {title} {Scaling {{Laws}} of
  {{Collective Ride-Sharing Dynamics}}},\ }\href
  {https://doi.org/10.1103/PhysRevLett.125.248302} {\bibfield  {journal}
  {\bibinfo  {journal} {Phys. Rev. Lett.}\ }\textbf {\bibinfo {volume} {125}},\
  \bibinfo {pages} {248302} (\bibinfo {year} {2020})}\BibitemShut {NoStop}%
\bibitem [{\citenamefont {Zech}\ \emph {et~al.}(2022)\citenamefont {Zech},
  \citenamefont {Molkenthin}, \citenamefont {Timme},\ and\ \citenamefont
  {Schr{\"o}der}}]{zech2022}%
  \BibitemOpen
  \bibfield  {author} {\bibinfo {author} {\bibfnamefont {R.~M.}\ \bibnamefont
  {Zech}}, \bibinfo {author} {\bibfnamefont {N.}~\bibnamefont {Molkenthin}},
  \bibinfo {author} {\bibfnamefont {M.}~\bibnamefont {Timme}},\ and\ \bibinfo
  {author} {\bibfnamefont {M.}~\bibnamefont {Schr{\"o}der}},\ }\bibfield
  {title} {\bibinfo {title} {Collective dynamics of capacity-constrained
  ride-pooling fleets},\ }\href {https://doi.org/10.1038/s41598-022-14960-x}
  {\bibfield  {journal} {\bibinfo  {journal} {Sci. Rep.}\ }\textbf {\bibinfo
  {volume} {12}},\ \bibinfo {pages} {10880} (\bibinfo {year}
  {2022})}\BibitemShut {NoStop}%
\bibitem [{\citenamefont {Agatz}\ \emph {et~al.}(2012)\citenamefont {Agatz},
  \citenamefont {Erera}, \citenamefont {Savelsbergh},\ and\ \citenamefont
  {Wang}}]{agatz2012}%
  \BibitemOpen
  \bibfield  {author} {\bibinfo {author} {\bibfnamefont {N.}~\bibnamefont
  {Agatz}}, \bibinfo {author} {\bibfnamefont {A.}~\bibnamefont {Erera}},
  \bibinfo {author} {\bibfnamefont {M.}~\bibnamefont {Savelsbergh}},\ and\
  \bibinfo {author} {\bibfnamefont {X.}~\bibnamefont {Wang}},\ }\bibfield
  {title} {\bibinfo {title} {Optimization for {{Dynamic Ride-Sharing}}: {{A
  Review}}},\ }\href {https://doi.org/10.1016/j.ejor.2012.05.028} {\bibfield
  {journal} {\bibinfo  {journal} {Eur. J. Oper. Res.}\ }\textbf {\bibinfo
  {volume} {223}},\ \bibinfo {pages} {295} (\bibinfo {year}
  {2012})}\BibitemShut {NoStop}%
\bibitem [{\citenamefont {Vazifeh}\ \emph {et~al.}(2018)\citenamefont
  {Vazifeh}, \citenamefont {Santi}, \citenamefont {Resta}, \citenamefont
  {Strogatz},\ and\ \citenamefont {Ratti}}]{vazifeh2018}%
  \BibitemOpen
  \bibfield  {author} {\bibinfo {author} {\bibfnamefont {M.~M.}\ \bibnamefont
  {Vazifeh}}, \bibinfo {author} {\bibfnamefont {P.}~\bibnamefont {Santi}},
  \bibinfo {author} {\bibfnamefont {G.}~\bibnamefont {Resta}}, \bibinfo
  {author} {\bibfnamefont {S.~H.}\ \bibnamefont {Strogatz}},\ and\ \bibinfo
  {author} {\bibfnamefont {C.}~\bibnamefont {Ratti}},\ }\bibfield  {title}
  {\bibinfo {title} {Addressing the {{Minimum Fleet Problem}} in {{On-Demand
  Urban Mobility}}},\ }\href {https://doi.org/10.1038/s41586-018-0095-1}
  {\bibfield  {journal} {\bibinfo  {journal} {Nature}\ }\textbf {\bibinfo
  {volume} {557}},\ \bibinfo {pages} {534} (\bibinfo {year}
  {2018})}\BibitemShut {NoStop}%
\bibitem [{\citenamefont {Spieser}\ \emph {et~al.}(2014)\citenamefont
  {Spieser}, \citenamefont {Treleaven}, \citenamefont {Zhang}, \citenamefont
  {Frazzoli}, \citenamefont {Morton},\ and\ \citenamefont
  {Pavone}}]{spieser2014}%
  \BibitemOpen
  \bibfield  {author} {\bibinfo {author} {\bibfnamefont {K.}~\bibnamefont
  {Spieser}}, \bibinfo {author} {\bibfnamefont {K.}~\bibnamefont {Treleaven}},
  \bibinfo {author} {\bibfnamefont {R.}~\bibnamefont {Zhang}}, \bibinfo
  {author} {\bibfnamefont {E.}~\bibnamefont {Frazzoli}}, \bibinfo {author}
  {\bibfnamefont {D.}~\bibnamefont {Morton}},\ and\ \bibinfo {author}
  {\bibfnamefont {M.}~\bibnamefont {Pavone}},\ }\bibfield  {title} {\bibinfo
  {title} {Toward a {{Systematic Approach}} to the {{Design}} and
  {{Evaluation}} of {{Automated Mobility-on-Demand Systems}}: {{A Case Study}}
  in {{Singapore}}},\ }in\ \href
  {https://doi.org/10.1007/978-3-319-05990-7\_20} {\emph {\bibinfo {booktitle}
  {Road {{Vehicle Automation}}}}},\ \bibinfo {editor} {edited by\ \bibinfo
  {editor} {\bibfnamefont {G.}~\bibnamefont {Meyer}}\ and\ \bibinfo {editor}
  {\bibfnamefont {S.}~\bibnamefont {Beiker}}}\ (\bibinfo  {publisher}
  {{Springer International Publishing}},\ \bibinfo {address} {{Cham}},\
  \bibinfo {year} {2014})\ pp.\ \bibinfo {pages} {229--245}\BibitemShut
  {NoStop}%
\bibitem [{\citenamefont {Zwick}\ \emph
  {et~al.}(2021{\natexlab{b}})\citenamefont {Zwick}, \citenamefont {Kuehnel},
  \citenamefont {Moeckel},\ and\ \citenamefont {Axhausen}}]{zwick2021agent}%
  \BibitemOpen
  \bibfield  {author} {\bibinfo {author} {\bibfnamefont {F.}~\bibnamefont
  {Zwick}}, \bibinfo {author} {\bibfnamefont {N.}~\bibnamefont {Kuehnel}},
  \bibinfo {author} {\bibfnamefont {R.}~\bibnamefont {Moeckel}},\ and\ \bibinfo
  {author} {\bibfnamefont {K.~W.}\ \bibnamefont {Axhausen}},\ }\bibfield
  {title} {\bibinfo {title} {Agent-based simulation of city-wide autonomous
  ride-pooling and the impact on traffic noise},\ }\href
  {https://doi.org/10.1016/j.trd.2020.102673} {\bibfield  {journal} {\bibinfo
  {journal} {Transp. Res. D}\ }\textbf {\bibinfo {volume} {90}},\ \bibinfo
  {pages} {102673} (\bibinfo {year} {2021}{\natexlab{b}})}\BibitemShut
  {NoStop}%
\bibitem [{\citenamefont {Barann}\ \emph {et~al.}(2017)\citenamefont {Barann},
  \citenamefont {Beverungen},\ and\ \citenamefont {M{\"u}ller}}]{barann2017}%
  \BibitemOpen
  \bibfield  {author} {\bibinfo {author} {\bibfnamefont {B.}~\bibnamefont
  {Barann}}, \bibinfo {author} {\bibfnamefont {D.}~\bibnamefont {Beverungen}},\
  and\ \bibinfo {author} {\bibfnamefont {O.}~\bibnamefont {M{\"u}ller}},\
  }\bibfield  {title} {\bibinfo {title} {An {{Open-Data Approach}} for
  {{Quantifying}} the {{Potential}} of {{Taxi Ridesharing}}},\ }\href
  {https://doi.org/10.1016/j.dss.2017.05.008} {\bibfield  {journal} {\bibinfo
  {journal} {Decis. Support. Syst.}\ }\textbf {\bibinfo {volume} {99}},\
  \bibinfo {pages} {86} (\bibinfo {year} {2017})}\BibitemShut {NoStop}%
\bibitem [{\citenamefont {Herminghaus}(2019)}]{herminghaus2019}%
  \BibitemOpen
  \bibfield  {author} {\bibinfo {author} {\bibfnamefont {S.}~\bibnamefont
  {Herminghaus}},\ }\bibfield  {title} {\bibinfo {title} {Mean {{Field Theory}}
  of {{Demand Responsive Ride Pooling Systems}}},\ }\href
  {https://doi.org/10.1016/j.tra.2018.10.028} {\bibfield  {journal} {\bibinfo
  {journal} {Transp. Res. A}\ }\textbf {\bibinfo {volume} {119}},\ \bibinfo
  {pages} {15} (\bibinfo {year} {2019})}\BibitemShut {NoStop}%
\bibitem [{\citenamefont {Manik}\ and\ \citenamefont
  {Molkenthin}(2020)}]{manik2020}%
  \BibitemOpen
  \bibfield  {author} {\bibinfo {author} {\bibfnamefont {D.}~\bibnamefont
  {Manik}}\ and\ \bibinfo {author} {\bibfnamefont {N.}~\bibnamefont
  {Molkenthin}},\ }\bibfield  {title} {\bibinfo {title} {Topology dependence of
  on-demand ride-sharing},\ }\href {https://doi.org/10.1007/s41109-020-00290-2}
  {\bibfield  {journal} {\bibinfo  {journal} {Appl. Netw. Sci.}\ }\textbf
  {\bibinfo {volume} {5}},\ \bibinfo {pages} {49} (\bibinfo {year}
  {2020})}\BibitemShut {NoStop}%
\bibitem [{\citenamefont {Fielbaum}(2021)}]{fielbaum2021b}%
  \BibitemOpen
  \bibfield  {author} {\bibinfo {author} {\bibfnamefont {A.}~\bibnamefont
  {Fielbaum}},\ }\bibfield  {title} {\bibinfo {title} {Optimizing a
  {{Vehicle}}'s {{Route}} in an on-{{Demand Ridesharing System}} in {{Which
  Users Might Walk}}},\ }\href {https://doi.org/10.1080/15472450.2021.1901225}
  {\bibfield  {journal} {\bibinfo  {journal} {J. Intell. Transp. Syst.}\
  }\textbf {\bibinfo {volume} {26}},\ \bibinfo {pages} {32} (\bibinfo {year}
  {2021})}\BibitemShut {NoStop}%
\bibitem [{\citenamefont {{de Ruijter}}\ \emph {et~al.}(2020)\citenamefont {{de
  Ruijter}}, \citenamefont {Cats}, \citenamefont {{Alonso-Mora}},\ and\
  \citenamefont {Hoogendoorn}}]{deruijter2020}%
  \BibitemOpen
  \bibfield  {author} {\bibinfo {author} {\bibfnamefont {A.}~\bibnamefont {{de
  Ruijter}}}, \bibinfo {author} {\bibfnamefont {O.}~\bibnamefont {Cats}},
  \bibinfo {author} {\bibfnamefont {J.}~\bibnamefont {{Alonso-Mora}}},\ and\
  \bibinfo {author} {\bibfnamefont {S.}~\bibnamefont {Hoogendoorn}},\
  }\bibfield  {title} {\bibinfo {title} {Ride-sharing efficiency and level of
  service under alternative demand, behavioral and pricing settings},\ }in\
  \href@noop {} {\emph {\bibinfo {booktitle} {Transportation {{Research Board}}
  2020 {{Annual Meeting}}}}}\ (\bibinfo {year} {2020})\BibitemShut {NoStop}%
\bibitem [{\citenamefont {de~Ruijter}\ \emph {et~al.}(2023)\citenamefont
  {de~Ruijter}, \citenamefont {Cats}, \citenamefont {Alonso-Mora},\ and\
  \citenamefont {Hoogendoorn}}]{deRuijter2023ride}%
  \BibitemOpen
  \bibfield  {author} {\bibinfo {author} {\bibfnamefont {A.}~\bibnamefont
  {de~Ruijter}}, \bibinfo {author} {\bibfnamefont {O.}~\bibnamefont {Cats}},
  \bibinfo {author} {\bibfnamefont {J.}~\bibnamefont {Alonso-Mora}},\ and\
  \bibinfo {author} {\bibfnamefont {S.}~\bibnamefont {Hoogendoorn}},\
  }\bibfield  {title} {\bibinfo {title} {Ride-pooling adoption, efficiency and
  level of service under alternative demand, behavioural and pricing
  settings},\ }\href {https://doi.org/10.1080/03081060.2023.2194874} {\bibfield
   {journal} {\bibinfo  {journal} {Transp. Plan. Technol.}\ }\textbf {\bibinfo
  {volume} {46}},\ \bibinfo {pages} {407} (\bibinfo {year} {2023})}\BibitemShut
  {NoStop}%
\bibitem [{\citenamefont {M{\"u}hle}(2023)}]{muhle2023}%
  \BibitemOpen
  \bibfield  {author} {\bibinfo {author} {\bibfnamefont {S.}~\bibnamefont
  {M{\"u}hle}},\ }\bibfield  {title} {\bibinfo {title} {An analytical framework
  for modeling ride pooling efficiency and minimum fleet size},\ }\href
  {https://doi.org/10.1016/j.multra.2023.100080} {\bibfield  {journal}
  {\bibinfo  {journal} {Multimodal Transp.}\ }\textbf {\bibinfo {volume} {2}},\
  \bibinfo {pages} {100080} (\bibinfo {year} {2023})}\BibitemShut {NoStop}%
\bibitem [{\citenamefont {Beojone}\ and\ \citenamefont
  {Geroliminis}(2021)}]{beojone2021}%
  \BibitemOpen
  \bibfield  {author} {\bibinfo {author} {\bibfnamefont {C.~V.}\ \bibnamefont
  {Beojone}}\ and\ \bibinfo {author} {\bibfnamefont {N.}~\bibnamefont
  {Geroliminis}},\ }\bibfield  {title} {\bibinfo {title} {On the
  {{Inefficiency}} of {{Ride-Sourcing Services}} towards {{Urban
  Congestion}}},\ }\href {https://doi.org/10.1016/j.trc.2020.102890} {\bibfield
   {journal} {\bibinfo  {journal} {Transp. Res. C}\ }\textbf {\bibinfo {volume}
  {124}},\ \bibinfo {pages} {102890} (\bibinfo {year} {2021})}\BibitemShut
  {NoStop}%
\bibitem [{\citenamefont {Engelhardt}\ \emph {et~al.}(2019)\citenamefont
  {Engelhardt}, \citenamefont {Dandl}, \citenamefont {Bilali},\ and\
  \citenamefont {Bogenberger}}]{engelhardt2019}%
  \BibitemOpen
  \bibfield  {author} {\bibinfo {author} {\bibfnamefont {R.}~\bibnamefont
  {Engelhardt}}, \bibinfo {author} {\bibfnamefont {F.}~\bibnamefont {Dandl}},
  \bibinfo {author} {\bibfnamefont {A.}~\bibnamefont {Bilali}},\ and\ \bibinfo
  {author} {\bibfnamefont {K.}~\bibnamefont {Bogenberger}},\ }\bibfield
  {title} {\bibinfo {title} {Quantifying the {{Benefits}} of {{Autonomous
  On-Demand Ride-Pooling}}: {{A Simulation Study}} for {{Munich}},
  {{Germany}}},\ }in\ \href {https://doi.org/10.1109/ITSC.2019.8916955} {\emph
  {\bibinfo {booktitle} {The 2019 {{IEEE Intelligent Transportation Systems
  Conference}} - {{ITSC}}}}}\ (\bibinfo  {publisher} {{IEEE}},\ \bibinfo {year}
  {2019})\ pp.\ \bibinfo {pages} {2992--2997}\BibitemShut {NoStop}%
\bibitem [{\citenamefont {Fagnant}\ and\ \citenamefont
  {Kockelman}(2018)}]{fagnant2018}%
  \BibitemOpen
  \bibfield  {author} {\bibinfo {author} {\bibfnamefont {D.~J.}\ \bibnamefont
  {Fagnant}}\ and\ \bibinfo {author} {\bibfnamefont {K.~M.}\ \bibnamefont
  {Kockelman}},\ }\bibfield  {title} {\bibinfo {title} {Dynamic
  {{Ride-Sharing}} and {{Fleet Sizing}} for a {{System}} of {{Shared Autonomous
  Vehicles}} in {{Austin}}, {{Texas}}},\ }\href
  {https://doi.org/10.1007/s11116-016-9729-z} {\bibfield  {journal} {\bibinfo
  {journal} {Transportation}\ }\textbf {\bibinfo {volume} {45}},\ \bibinfo
  {pages} {143} (\bibinfo {year} {2018})}\BibitemShut {NoStop}%
\bibitem [{\citenamefont {Ruch}\ \emph {et~al.}(2020)\citenamefont {Ruch},
  \citenamefont {Lu}, \citenamefont {Sieber},\ and\ \citenamefont
  {Frazzoli}}]{ruch2020}%
  \BibitemOpen
  \bibfield  {author} {\bibinfo {author} {\bibfnamefont {C.}~\bibnamefont
  {Ruch}}, \bibinfo {author} {\bibfnamefont {C.}~\bibnamefont {Lu}}, \bibinfo
  {author} {\bibfnamefont {L.}~\bibnamefont {Sieber}},\ and\ \bibinfo {author}
  {\bibfnamefont {E.}~\bibnamefont {Frazzoli}},\ }\bibfield  {title} {\bibinfo
  {title} {Quantifying the {{Efficiency}} of {{Ride Sharing}}},\ }\href
  {https://doi.org/10.1109/TITS.2020.2990202} {\bibfield  {journal} {\bibinfo
  {journal} {IEEE Trans. Intell. Transp. Syst.}\ }\textbf {\bibinfo {volume}
  {22}},\ \bibinfo {pages} {5811 } (\bibinfo {year} {2020})}\BibitemShut
  {NoStop}%
\bibitem [{\citenamefont {Engelhardt}\ \emph {et~al.}(2020)\citenamefont
  {Engelhardt}, \citenamefont {Dandl},\ and\ \citenamefont
  {Bogenberger}}]{engelhardt2020}%
  \BibitemOpen
  \bibfield  {author} {\bibinfo {author} {\bibfnamefont {R.}~\bibnamefont
  {Engelhardt}}, \bibinfo {author} {\bibfnamefont {F.}~\bibnamefont {Dandl}},\
  and\ \bibinfo {author} {\bibfnamefont {K.}~\bibnamefont {Bogenberger}},\
  }\bibfield  {title} {\bibinfo {title} {Speed-up {{Heuristic}} for an
  {{On-Demand Ride-Pooling Algorithm}}},\ }\bibfield  {journal} {\bibinfo
  {journal} {ArXiv preprint}\ }\href
  {https://doi.org/10.48550/arXiv.2007.14877} {10.48550/arXiv.2007.14877}
  (\bibinfo {year} {2020})\BibitemShut {NoStop}%
\bibitem [{\citenamefont {Ma}\ and\ \citenamefont {{and
  Wolfson}}(2013)}]{ma2013a}%
  \BibitemOpen
  \bibfield  {author} {\bibinfo {author} {\bibfnamefont {S.}~\bibnamefont
  {Ma}}\ and\ \bibinfo {author} {\bibfnamefont {O.}~\bibnamefont {{and
  Wolfson}}},\ }\bibfield  {title} {\bibinfo {title} {T-{{Share}}: {{A
  Large-Scale Dynamic Taxi Ridesharing Service}}}\ }(\bibinfo  {publisher}
  {{ICDE 2013}},\ \bibinfo {year} {2013})\BibitemShut {NoStop}%
\bibitem [{\citenamefont {Lotze}\ \emph {et~al.}(2022)\citenamefont {Lotze},
  \citenamefont {Marszal}, \citenamefont {Schr{\"o}der},\ and\ \citenamefont
  {Timme}}]{lotze2022}%
  \BibitemOpen
  \bibfield  {author} {\bibinfo {author} {\bibfnamefont {C.}~\bibnamefont
  {Lotze}}, \bibinfo {author} {\bibfnamefont {P.}~\bibnamefont {Marszal}},
  \bibinfo {author} {\bibfnamefont {M.}~\bibnamefont {Schr{\"o}der}},\ and\
  \bibinfo {author} {\bibfnamefont {M.}~\bibnamefont {Timme}},\ }\bibfield
  {title} {\bibinfo {title} {Dynamic {{Stop Pooling}} for {{Flexible}} and
  {{Sustainable Ride Sharing}}},\ }\href
  {https://doi.org/10.1088/1367-2630/ac47c9} {\bibfield  {journal} {\bibinfo
  {journal} {New J. Phys.}\ }\textbf {\bibinfo {volume} {24}},\ \bibinfo
  {pages} {023034} (\bibinfo {year} {2022})}\BibitemShut {NoStop}%
\bibitem [{\citenamefont {Kersting}\ \emph {et~al.}(2021)\citenamefont
  {Kersting}, \citenamefont {Kallbach},\ and\ \citenamefont
  {Schl{\"u}ter}}]{kersting2021young}%
  \BibitemOpen
  \bibfield  {author} {\bibinfo {author} {\bibfnamefont {M.}~\bibnamefont
  {Kersting}}, \bibinfo {author} {\bibfnamefont {F.}~\bibnamefont {Kallbach}},\
  and\ \bibinfo {author} {\bibfnamefont {J.~C.}\ \bibnamefont {Schl{\"u}ter}},\
  }\bibfield  {title} {\bibinfo {title} {For the young and old alike--an
  analysis of the determinants of seniors’ satisfaction with the true
  door-to-door {{DRT}} system {{EcoBus}} in rural {{Germany}}},\ }\href
  {https://doi.org/10.1016/j.jtrangeo.2021.103173} {\bibfield  {journal}
  {\bibinfo  {journal} {J. Transp. Geogr.}\ }\textbf {\bibinfo {volume} {96}},\
  \bibinfo {pages} {103173} (\bibinfo {year} {2021})}\BibitemShut {NoStop}%
\bibitem [{\citenamefont {S{\"o}rensen}\ \emph {et~al.}(2021)\citenamefont
  {S{\"o}rensen}, \citenamefont {Bossert}, \citenamefont {Jokinen},\ and\
  \citenamefont {Schl{\"u}ter}}]{sorensen2021much}%
  \BibitemOpen
  \bibfield  {author} {\bibinfo {author} {\bibfnamefont {L.}~\bibnamefont
  {S{\"o}rensen}}, \bibinfo {author} {\bibfnamefont {A.}~\bibnamefont
  {Bossert}}, \bibinfo {author} {\bibfnamefont {J.-P.}\ \bibnamefont
  {Jokinen}},\ and\ \bibinfo {author} {\bibfnamefont {J.}~\bibnamefont
  {Schl{\"u}ter}},\ }\bibfield  {title} {\bibinfo {title} {How much flexibility
  does rural public transport need?--{{Implications}} from a fully flexible
  {{DRT}} system},\ }\href {https://doi.org/10.1016/j.tranpol.2020.09.005}
  {\bibfield  {journal} {\bibinfo  {journal} {Transp. Policy}\ }\textbf
  {\bibinfo {volume} {100}},\ \bibinfo {pages} {5} (\bibinfo {year}
  {2021})}\BibitemShut {NoStop}%
\bibitem [{\citenamefont {{OpenStreetMap contributors}}(2021)}]{OSM}%
  \BibitemOpen
  \bibfield  {author} {\bibinfo {author} {\bibnamefont {{OpenStreetMap
  contributors}}},\ }\href {https://www.openstreetmap.org} {\bibinfo {title}
  {{{OpenStreetMap}}}} (\bibinfo {year} {2021})\BibitemShut {NoStop}%
\bibitem [{\citenamefont {Erhardt}(2021)}]{erhardt2021a}%
  \BibitemOpen
  \bibfield  {author} {\bibinfo {author} {\bibfnamefont {S.}~\bibnamefont
  {Erhardt}},\ }\href {https://opentopomap.org/} {\bibinfo {title}
  {{{OpenTopoMap}}}} (\bibinfo {year} {2021})\BibitemShut {NoStop}%
\bibitem [{\citenamefont {{OSRM contributors}}(2022)}]{osrm}%
  \BibitemOpen
  \bibfield  {author} {\bibinfo {author} {\bibnamefont {{OSRM contributors}}},\
  }\href {https://project-osrm.org/} {\bibinfo {title} {{{Open Source Routing
  Machine}}}} (\bibinfo {year} {2022})\BibitemShut {NoStop}%
\end{thebibliography}

%

\end{document}